\def\mev{\,{\rm Me\kern-0.1em V}}
\def\gev{\,{\rm Ge\kern-0.1em V}}
\def\alphacont{\alpha_{\sl cont}}
\def\alphalat{\alpha_{\sl lat}}
\def\Zcont{Z_{\sl cont}}
\def\mbstar{m_b^*}
\def\mbpole{m_{b\,\sl pole}}
\def\qstar{q^*}
\def\kappac{\kappa_{\sl c}}
\def\kappaconeloop{\kappa_{\sl c,\, one-loop}}

\def\vev#1{\langle #1\rangle}
\def\zerohat{\hat0}
\def\third{{\textstyle 1\over3}}
\documentstyle[11pt]{article}
\renewcommand{\baselinestretch}{1.5}
\begin{document}
\vspace*{-1.25in}
\small{
\begin{flushright}
FERMILAB-PUB-94/164-T \\[-.1in]
UCLA/94/TEP/3 \\[-.1in]
UVA-INPP-94-01 \\[-.1in]
Pitt-THY-94-08 \\[-.1in]
SHEP-93/94-28 \\[-.1in]
July, 1994 \\
\end{flushright}}
\vspace*{.85in}
\begin{center}
{\Large{\bf Properties of $B$-Mesons in Lattice QCD}} \\
\vspace*{.45in}
{\large{A.~Duncan$^1$, E.~Eichten$^2$, J.~Flynn$^3$, B.~Hill$^4$,
G.~Hockney$^2$, and
H.~Thacker$^5$}} \\
\vspace*{.15in}
$^1$Dept. of Physics and Astronomy, Univ. of Pittsburgh, Pittsburgh, PA 15620\\
$^2$Fermilab, P.O. Box 500, Batavia, IL 60510 \\
$^3$Physics Dept., Univ. of Southampton, Southampton SO17 1BJ UK \\
$^4$Dept. of Physics, Univ. of California, Los Angeles, CA 90024\\
$^5$Dept.of Physics, University of Virginia, Charlottesville, VA 22901
\end{center}
\vspace*{.3in}
\begin{abstract}
The results of an extensive study of B-meson properties in quenched lattice
QCD are presented. The studies are carried out in the static quark limit
where the b-quark is taken to be infinitely massive. Our computations rely
on a multistate smearing method introduced previously,
with smearing functions generated from a relativistic lattice quark model.
Systematic errors arising from
excited state contamination, finite volume effects, and the chiral
extrapolation
for the light quarks are estimated.
We obtain continuum results for the mass splitting
$M_{B_s}-M_{B_u} = 86 \pm 12(stat) ~^{+7}_{-9}(syst)$ \mev,
the ratio of decay constants
$f_{B_s}/f_{B_u} = 1.22 \pm 0.04(stat)
\pm 0.02 (syst)$.
For the B-meson decay constant
we separately exhibit the sizable
uncertainties in the extrapolation
to the continuum limit ($a\rightarrow 0$)
and higher order perturbative matching.
We obtain
$f_{B} = 188 \pm 23(stat) \pm 15(syst)~^{+26}_{-0}(extrap) \pm 14 (pert)$
\mev .
\end{abstract}


\newpage
\section{INTRODUCTION}


Heavy-light mesons are ideal systems for lattice QCD studies.
For hadrons that contain a single heavy quark, the
dynamics of QCD simplifies as $m_Q \rightarrow \infty$.
In this limit, the heavy quark is on shell and at rest
relative to the hadron\cite{Eic81} and the
QCD dynamics becomes independent of $m_Q$\cite{Proof90}.
Heavy quark mass dependence can be extracted
analytically to produce an effective action for a static
quark and the remaining
light degrees of freedom\cite{EichtenHill1,Georgi90}.
In the continuum, the resulting heavy quark
effective theory (HQET) makes transparent the
symmetry\cite{Wisgur} and scaling relations
between systems which differ
by heavy quark spin or flavor \cite{Wisgur,Falk,Wise92}.
Furthermore for finite heavy quark masses,
the effective action can be improved order by order in $1/m_Q$.
The $m_Q \rightarrow\infty$ limit
may also provide other insights into
QCD dynamics.  In particular, heavy-light mesons are physical systems
with a single light valence quark and fully relativistic QCD dynamics.
It is likely that much can be learned about constituent
quark ideas in this simple setting\cite{Bj}.

These theoretical developments
have immediate physical applications for B physics.
Since the b quark mass is
significantly heavier than the other mass scales (the QCD scale and the
light quarks masses) which enter into the dynamics of B hadrons, it
is likely to be a good approximation to treat the b quark
in the $m_Q \rightarrow \infty$ limit within B hadrons.

Recent developments in lattice gauge theory have led to the
possibility of calculating the masses and decay
constants of $B_q$ ($q$=u, d, s)
mesons from first principles (QCD) with enough accuracy to be of
both phenomenological and theoretical interest.
In particular, we consider the mass difference $M_{B_s} - M_{B_u}$,
the decay constant $f_{B_{u,d}}$
and the ratio $f_{B_s}/f_{B_d}$.
We will only consider the lattice action appropriate to
the static limit ($m_b \rightarrow \infty$) and hence all the results
reported have corrections of order $\Lambda_{QCD}/m_b$.
A variety of other methods have been developed to study B mesons on the
lattice.  Methods for treating the b quark
using nonrelativistic actions\cite{NR90},
a non-zero velocity formulation\cite{Mandula},
and a generalization of the usual Wilson action which is not
constrained to quark masses less than the inverse lattice
spacing\cite{KronMack} are being actively pursued.

  The present study encompasses a systematic analysis of data for
$M_B$ and $f_B$ at four
different lattice spacings $a$
(with associated beta values $\beta = 5.7, 5.9, 6.1$, and $6.3$)
and a
variety of physical volumes (in lattice units $12^3$, $16^3$, and $20^3$)
for one fixed spacing ($\beta = 5.9$).
For each case above at least four light quark mass values (kappa values
corresponding to pion masses in the range 300 - 800 MeV)
were studied.
A uniform and consistent fitting scheme was used in all cases.
This is particularly important
for an accurate extrapolation to results at  physical light
quark masses ($\kappa= \kappa_{u,d}$ and $\kappa= \kappa_{s}$).
and for assessing the $a$-dependence of
the results.

In the static approximation, the heavy quark
propagator is reduced to a straight timelike Wilson line, making it possible
to calculate correlators of spatially smeared $\bar{Q}q$ operators without
having to compute light quark propagators from smeared sources. Because of
this simplifying feature, the heavy-light meson system is an ideal place
to develop sophisticated operator smearing techniques. Such techniques
are indispensable for the accurate calculation of $f_B$ and other B-meson
properties. Until recently, most such calculations have relied on a
more-or-less
ad hoc choice of smearing functions (e.g. walls\cite{BLS},
cubes \cite{cubes}, or exponentials \cite{Wuppertal},
or Gaussians \cite{Wuppertal}).

In the present study, we have applied a multistate smearing method introduced
previously in Ref. \cite{lat91_multistate}. This analysis provides some
significant improvements over previous investigations.
First, we have made a serious effort to construct
smearing functions which closely resemble the actual Coulomb gauge wave
functions of the valence $\bar{Q}q$ system as measured on the lattice. As
reported in a previous paper \cite{DET_RQM}, the heavy-light wavefunctions from
lattice QCD are reproduced with remarkable accuracy by a simple relativistic
quark model (RQM) Hamiltonian which contains the static QCD potential extracted
from Wilson lines in Coulomb gauge. In addition to being an interesting
statement about QCD dynamics, the success of the RQM has a practical
consequence
which we will exploit here. The RQM Hamiltonian provides a simple and
precise way of constructing orthonormal sets of realistic smearing functions
for lattice
heavy-light calculations. For each value of $\beta$ and lattice size, the
static Wilson potential is calculated from the gauge configurations and used
in the RQM to generate heavy-light smearing functions. The only tuneable
parameter in this procedure is the light-quark constituent mass parameter $\mu$
in the
kinetic term of the RQM Hamiltonian. In practice, this parameter was initially
selected by measuring the lattice QCD ground state wavefunction and adjusting
$\mu$ to give the best fit for the RQM ground state. In some cases, after an
initial multistate fit to the heavy-light propagators, it was found useful
to iterate the procedure with a more finely tuned value of the constituent
quark mass $\mu$, using the
more accurate wave functions obtained from the multistate fit.

In addition to this method for constructing smearing functions, another
important innovation introduced in the present study is the fitting procedure
used to extract information from the heavy-light correlators. Starting with
the wavefunctions from the first M S-wave states of the RQM, it is relatively
easy to construct the entire $M\times M$ matrix of correlators among the
corresponding smeared $\bar{Q}q$ operators, as well as the
``smeared-local'' correlators between each of these operators and the local
$\bar{Q}q$ source. Along with the local-local correlator, these form an
$(M+1)\times (M+1)$ matrix. This matrix contains far more information than
just the smeared-smeared and smeared-local correlators of any single smearing
function. In particular, the matrix contains information about excited states,
which, when properly exploited, allows an accurate extraction of ground state
properties even at very short time separations, where excited state
contributions
are still large. The method we introduce to accomplish this employs a $\chi^2$
minimization procedure to simultaneously fit the $(M+1)\times (M+1)$
matrix of correlators
to a sum of $M$ exponential (pole) terms, representing
the contribution of the $M$ lowest lying heavy-light eigenstates. (In all fits,
we exclude the local-local correlator, which, at short time separations,
is not well fit by a few low lying states.)
The matrix
coefficient (residue) of each pole term factorizes and can be written in
terms of an $M+1$ component vector whose entries represent the
vacuum-to-eigenstate
matrix element of each smeared operator. In practice we have found an $M=2$
fit to yield fairly accurate results for ground state properties.
For all of the fits used, the $\chi^2$ per degree of freedom
was less than 1.3. The multiparameter fits were carried out using the
CERNLIB  minimization routine MINUIT.

The improved control over systematic errors
gained from the multistate fitting method
allows us to better address a number of issues.
In particular the dependence of heavy-light meson parameters on
both the light quark mass and the lattice spacing are examined in detail.
One of the difficulties with
previous analyses which prevented accurate chiral and $a\rightarrow 0$
extrapolations
was in the arbitrariness of the smearing procedure. It is clear that any
ad~hoc smearing function will have a substantial overlap with excited states.
Typically one tries to optimize the smearing function (e.g. by adjusting
the size of the cube) and to go far enough out in time that excited states
have died away. The approximate equality of smeared-smeared and
smeared-local effective masses, combined with some indication of an effective
mass plateau, are the main criteria of
success in this procedure. Unfortunately, the
procedure is somewhat subjective and it is difficult to rule out large
systematic errors due to excited state contamination. An attempt to reduce
these errors by extracting results from larger time separations leads to a
rapid deterioration of statistics. Moreover, because of the well-known
signal-to-noise difficulty for the heavy-light propagator in the static
approximation, \cite{noise,lat91_multistate}, the problem of isolating the
ground
state becomes more difficult at smaller lattice spacing.
As a result, extrapolation to $a=0$ is particularly problematic.
Furthermore, use of any fixed smearing function at different light quark
masses introduces a significant systematic error in the
extracted kappa dependence.  This is important
in the determination of results for $f_{B_s}/f_{B_u}$
and $M_{B_s} - M_{B_u}$.
The multistate fitting procedure effectively
deals with these difficulties, greatly reducing our errors.

Extrapolating to the continuum limit ($a = 0$) we obtain the
ratio $f_{B_{s}}/f_{B_{u}} = 1.22 \pm .04{(\rm stat)} \pm .02{(\rm syst)}$
and $M_{B_{s}}-M_{B_{u}} = 86 \pm 12 {(\rm stat)} ~^{+7}_{-9}{(\rm syst)}\mev$
for the mass difference.
For these quantities, only a slight dependence on the lattice spacing is
observed, and the systematic errors associated with the $a\rightarrow 0$
extrapolation (included in the above) are small.

The situation for the decay constant is more complicated.
We find a significant lattice spacing dependence for the
ground-state pseudoscalar decay constant $f_B$.
The results for the four $\beta$
values are consistent with either a linear or quadratic dependence
on the lattice spacing $a$.
The linearly extrapolated result in the $a\rightarrow 0$ limit is
$f_B = 188\pm 23\pm 15 \mev$
This result is notably smaller than previous
estimates of $f_B$ in the static approximation. The primary reason for this
is the $a\rightarrow 0$ extrapolation.
The quadratic extrapolated result in the $a\rightarrow 0$ limit is
$f_B = 214\pm 13\pm 17 \mev$.
This fit reflects the fact that our results at $\beta = 6.3$
are not inconsistent with those of Ref. \cite{BLS},\cite{Rome2},\cite{UKQCD}
(the latter two results using a clover action for the light quarks).
In both cases, the first error is statistical
while the second includes systematic errors, which we discuss in Section 5.
Our final result for the $B_{u,d}$ meson decay constant (in the
heavy quark limit)
$f_B = 188 \pm 23({\rm stat}) \pm 15({\rm syst})~^{+26}_{-0}({\rm extrap})
\pm 14({\rm pert}) \mev$
explicitly separates out this theoretical uncertainity
associated with the $a \rightarrow 0$ extrapolation as well
as our estimate of the uncertainity associated with higher
order perturbative matching corrections.

We will give a complete discussion of the perturbative
matching of lattice to continuum results in Section 2. We include
a discussion of the heavy quark mass renormalization in relation to
the residual mass parameter ($\bar\Lambda = M_B - m_b({\rm quark})$)
of HQET.
In Section 3 we discuss our analysis procedure. The details
of the multistate smearing technique
and the construction of the smearing functions
from a relativistic quark model (RQM) are presented.
In Section 4 the numerical lattice results at each $\beta$ are presented.
The statistical
and fitting errors associated with our final physical results
are determined. The light meson results used to set light quark
masses are contained in Appendix A.
The discussion of systematic errors associated
with excited state contamination, finite volume, nonzero light quark masses,
scale uncertainities, and the extrapolation to zero lattice spacing
are all discussed in Section 5.
A study of the time evolution of the wavefunctions for
heavy-light states is presented in Appendix B.
These results provide an independent check
that our multistate smearing analysis
has removed excited state contamination.
In the Section 6 we present our final results,
compare them with other recent calculations, and discuss
upcoming studies.


\section{Perturbative Matching}\label{sec:pert}%

\subsection{Extracting Properties of Heavy-Light Mesons
from LQCD}\label{sec2intro}%
In this section, we will focus on short-distance corrections
to the results obtained from lattice~QCD.  These
corrections are common to both traditional techniques for extracting
meson properties and to the multistate smearing method employed here.
Explanation of the details of the multistate smearing method are deferred
to the next section.

The corrections to the matrix elements of time component
of the heavy-light axial current, $J_{05}(n)$, are
computed by demanding that the ratio of the current renormalized with
some continuum regularization scheme and the lattice-regularized
current be unity.  To be a little more precise, one computes this
ratio,~$Z$, using some matrix element, and from then on one multiplies
any result obtained using the lattice-regularized current by~$Z$.  The
states used to determine~$Z$ can be chosen for calculational
convenience since the ratio is independent of the choice of states.
Although there is no choice of states for which the numerator and
denominator of the ratio are separately calculable, because the
operators only differ at scales on the order of the cutoffs of the two
regularizations, and at these scales QCD is perturbative, the ratio
can be calculated in perturbation theory.

The procedure is actually slightly more complicated than explained
in the preceding paragraph.  Because the lattice~QCD calculations
are done in the heavy quark effective theory (a theory which does
not have the same particle content as the
full standard model), it is necessary to compute an additional ratio, $\Zcont$.
This is the ratio of the axial current renormalized at the scale
$\mbstar$ in the standard model to the axial current renormalized at
a scale $\qstar$ in the heavy quark effective theory.

The calculations of $Z$ and $\Zcont$ are the subject of the following
two subsections. An analysis of the heavy quark mass
renormalization follows thereafter.
In the last subsection, we summarize the various constants
used in this study.

\subsection{Calculation of $Z$}%
The ratio $Z$ introduced above was calculated some time ago, but
there is substantial uncertainty in these calculations.  Tadpole-improved
perturbation theory, as formulated by Lepage and Mackenzie \cite{LepMack},
promises
to reduce these uncertainties below the ten per cent level at
one-loop.  The application of the tadpole-improvement program to heavy
quark effective theory has recently been discussed by Bernard \cite{bernard},
and calculations
have been performed by Hern\'andez and Hill \cite{HernHill}.
In this subsection we summarize the calculation of $Z$ within the
framework of tadpole-improved perturbation theory.
Hern\'andez and Hill considered both the zero-separation and
unit-separation point-split heavy-light axial currents.  We will restrict
our attention here to the zero-separation heavy-light current used in our
Monte Carlo calculations.  We will further restrict our attention
to the case of Wilson fermions with $r=1$.
The use of tadpole-improved perturbation
theory results in a substantial reduction in our best estimate of the
central value and uncertainty for $f_B$.

The root of the tadpole-improvement program is a nonperturbative
renormalization of the basic operators in the lattice
action.  These redefinitions absorb the large renormalizations arising
from lattice tadpole graphs.
A related additional part of the
Lepage-Mackenzie prescription is the use of a larger perturbative coupling.
If one uses $\beta$ to determine the
perturbative coupling, $\alphalat$, one-loop perturbative
corrections are consistently underestimated.  These perturbation
theory problems are due to the fact that $\alphalat$ is a
poor choice of expansion parameter.  For example at an inverse lattice
spacing of 2~GeV, the tadpole-improved expansion
parameter is $\alpha_V = 0.16$, which is twice as large as
$\alphalat$.  Lepage and Mackenzie argue that the
best way to arrive at $\alpha_V$ is from a non-perturbative lattice
determination of a perturbatively calculable quantity, such as the
gauge field plaquette expectation value.

Using tadpole-improvement of the Wilson action for quarks on the
lattice as a guide, one can perform tadpole-improvement of the
heavy quark action, and this has been done
in Ref.~\cite{bernard}.
Instead of discretizing
$$
\label{eq:scont}
S=\int d^4x \thinspace b^\dagger \left(i \partial_0 + g A_0\right) b
$$
as
$$
\label{eq:sdisc}
S=i a^3 \sum\nolimits_{n}
b^\dagger(n)
\left(
b(n)-\vphantom{1\over u_0}U_0(n{-}\zerohat)^\dagger b(n{-}\zerohat)
\right)\!,
$$
it is discretized as
$$
\label{eq:stad}
S_{\sl tadpole-improved}=i a^3 \sum\nolimits_{n}
b^\dagger(n)
\left(
b(n)-{1\over u_0}U_0(n{-}\zerohat)^\dagger b(n{-}\zerohat)
\right)\!,
$$
where $u_0$ is defined as
$$
u_0 \equiv \vev{\third {\rm Tr} U_{\sl plaq}}^{1/4}\!.
$$
The combination $U_\mu(x)/u_0$ more closely
corresponds to the continuum field $(1+iagA_\mu(x))$, than does
$U_\mu(x)$ itself.
With the tadpole-improved action, there is an additional factor of $1/u_0$ for
each gauge field link in the product.  Thus the Green's function of
two heavy-light currents separated by $n_0$ lattice spacings in the
time direction satisfies,
$$
[G_B(n_0)]_{\sl tadpole-improved} = {G_B(n_0) \over u_0^{n_0} }\!.
$$

The $B$ meson decay constant $f_B$ is usually extracted from numerical
simulations by fitting $G_B(n_0)$ to
$$
{(f_B m_B)^2\over 2 m_B} \exp[-Cn_0 a]
$$
Thus we see that the tadpole improvement procedure
has no effect on the fitted value of $f_B$.  Its only effect is the change
$$
C\rightarrow C+{\ln u_0\over a},
$$
that is, a linearly divergent mass renormalization.

So far we have seen that tadpole
improvement does not affect the extraction of $f_B$ as it
is generally done in lattice Monte Carlo calculations.
However we must still take into account
the effect of tadpole improvement of the light quark action, and this will
involve some additional factors.

As conventionally defined in lattice Monte Carlo calculations
the lattice operator $J^0_5$ involved in
calculating $f_B$ is renormalized by a factor
$\sqrt{2\kappac}Z$, where $\kappac$ is the critical value of the
hopping parameter for the light quarks.  The tree level value of
$\kappac$ is $1/8$.
Lepage and Mackenzie advocate a reorganization of perturbation
theory such that a factor of $\sqrt{8\kappac}$ is included in
$\tilde Z$ and the renormalizing factor becomes $\tilde Z/2$.

Let us see what this factor does at one-loop.
Calculations of $Z$ have been carried out to
one-loop order and the result is of the form,
$$
Z = 1 + {\alpha_S \over 3 \pi} [
\int {d^4 q\over \pi^2} g(q) + {3\over2}\ln(\qstar a)^2]\!,
$$
This definition of $g(q)$ (and similar ones for $h(q), j(q)$, and $k(q)$
which will be introduced below)
follow the definitions in Ref.~\cite{HernHill}.
A one-loop calculation of $8\kappac$ has been performed,
and the result is expressible as
$$
{1\over 8\kappaconeloop} = 1 - {\alpha_{\sl S}\over 3\pi }
\int {d^4q \over \pi^2} h(q)\!.
$$

The relationship between $\tilde Z$ and $Z$ is
$$
\tilde Z = \sqrt{8\kappaconeloop} Z \!.
$$
The one-loop expression for $\tilde Z$ is therefore
$$
\tilde g(q) = g(q) + h(q)/2\! ,
$$
where
$$
\tilde Z = 1 + {\alpha_S \over 3 \pi} [
\int {d^4 q\over \pi^2} \tilde g(q) + {3\over2}\ln(\qstar a)^2]\!.
$$

We continue with the application of the Lepage-Mackenzie
prescription to determine the $\Lambda$-value of the coupling and
the scale~$q^*$ at which it is evaluated.
The prescription for fixing the value of the coupling, $\alpha_V$
is to extract it from
a non-perturbative calculation of the
\hbox{1 $\times$ 1} Wilson loop ({\it i.e.,} the expectation value
of the plaquette, $U_{\sl plaq}$).  Once the coupling is known at some scale
(alternatively, once the value $\Lambda_V$ is known), it can be run
to any other scale.  The formula which relates $\alpha_V$
to the non-perturbatively determined (lattice
Monte Carlo) plaquette expectation value is,
$$
-\ln\vev{ \third {\rm Tr} U_{\sl plaq} }
= {4 \pi \over 3}  \alpha_V(3.41/a)
 \bigl[ 1- \alpha_V(3.41/a) (1.19 + 0.025 n_f) + O(\alpha_V^2) \bigr].
$$
The coefficient of $n_f$ is the one appropriate for Wilson fermions
with $r=1$.  In the quenched approximation, $n_f=0$.
$\Lambda_V$ is determined through
$$
\alpha_V(q) = \big[\beta_0 \ln(q^2/\Lambda_V^2)
               + (\beta_1/\beta_0) \ln\ln(q^2/\Lambda_V^2))\big]^{-1}\!.
$$

It remains to fix the scale $\qstar$ at which $\alpha_V(q)$ is evaluated in
the expression for $\tilde Z$.
Lepage and Mackenzie propose to do that by calculating the expectation value of
$\ln q^2$ in the one-loop perturbative lattice correction.  The formulae
determining this scale are:
\begin{eqnarray}
Y  &\equiv & \int d^4 q ~\tilde g(q) = -13.93 \\
\vev{\ln(qa)^2} &\equiv & {1\over Y} \int d^4 q ~\tilde g(q) \ln(q^2a^2)
= {21.76\over13.93}\\
q^*a &\equiv & \exp[\vev{\ln(qa)^2}/2] = 2.18\ .
\end{eqnarray}
Hernandez and Hill quote the errors on the numerically evaluated
values of $Y$ and $Y \ln<q^2>$
as order~1 in the last decimal place.

Using the two-loop formula for the
running of $\alpha_V$ with zero quark flavors
one obtains $\alpha_V(q^*)$.  The explicit
dependence on the value of $a^{-1}$ drops out of the ratio $q^*/\Lambda_V$.
Hence, the only way lattice Monte Carlo results have been used
so far is for the expectation value of the plaquette; the determination
of~$a^{-1}$ has not yet entered.
The results for~$Z$ at various values of~$\beta$
are summarized in Table~\ref{tab:scales}.

This completes our calculation of $\tilde Z$.
It remains to multiply~$\tilde Z$ by the continuum
running and matching factor,~$Z_{\sl cont}$.
We leave this for the following subsection.
\subsection{Continuum Running and Matching Factor}

For consistency, the one loop computation of $\tilde Z$ should be
combined with a two loop running in the continuum effective theory and
a further one loop matching between the continuum effective and full
theories.  This produces the continuum running and matching factor
$\Zcont$ which multiplies $\tilde Z$ to give the full perturbative
correction. It will turn out that the lattice to continuum matching
factor $\tilde Z$ is most significant, while $\Zcont$ produces only a
small additional change.

In the previous section,
the coupling $\alpha_V$ was determined in a no-flavor
(quenched) lattice theory.  We must now match onto a four-flavor (or
five-flavor depending on the value of $q^*$) continuum theory. In order for
the infrared behavior of lattice and continuum theories to match exactly, we
could use a continuum coupling whose value is equal to
$\alpha_V(q^*)$ (although differences between couplings are higher
order effects). This implies we should choose a continuum scale
$q_{\sl cont}$ according to $\alphacont(q_{\sl cont}) =
\alpha_V(q^*)$.
In practice we ignore this criterion and simply set $q_{\sl cont} =
q^*$, using $\alpha_V(q^*)$ everywhere in the matching. We then run in
the continuum theory using four or five flavors depending on whether
$q^*$ is greater or less than the $b$-quark threshold mass.

The $b$-quark threshold itself is determined as follows. We assume the
pole mass $\mbpole$ is known and relate it to the $\overline{\rm MS}$
running mass $m_b(\mu)$ according to~\cite{polemass}
\[
\mbpole = m_b(\mu) \left(
  1 + {\alphacont(\mu)\over\pi} \left[
     {4\over3} + \ln(\mu^2/m_b^2(\mu)) \right] \right).
\]
Setting $\mu = m_b^* = m_b(m_b^*)$, we fix the threshold mass $m_b^*$
by solving
\[
\mbpole = m_b(m_b^*) \left[
  1 + {4\alphacont(m_b^*)\over 3\pi} \right]
\]
We use the usual two loop result for $\alphacont$:
\def\moLsq{\mu^2/\Lambda^2}
\[
\alphacont = {4\pi\over b_0 \ln(\moLsq)} \left(
  1 - {b_1 \ln\ln(\moLsq)\over b_0 \ln(\moLsq)} \right)
\]
with
\[
b_0 = 11 - 2 n_f/3, \qquad b_1 = 102 - 38 n_f /3.
\]
Here $n_f$ is the number of light flavors and we take $\Lambda =
175\mev$ for five light flavors~\cite{pdg}.  For other $n_f$,
$\Lambda$ is fixed by demanding that $\alphacont$ be continuous.
Applying this procedure with $\mbpole = 4.72\gev$~\cite{polemassvalue}
gives
\[
m_b^* = 4.34\gev.
\]

Now that $q^*$ and $m_b^*$ are fixed we use them in the combined two
loop running plus one loop matching formula for $\Zcont$. The one-loop
anomalous dimension calculation~\cite{oneloopanomdim} for the
heavy-light axial current was extended to two loops by Ji and
Musolf~\cite{JandM}\ and the two loop result was confirmed by Broadhurst
and Grozin~\cite{BandGone}. The same authors~\cite{BandGtwo} also confirmed
the
one-loop matching calculation of Eichten and
Hill~\cite{EichtenHillI}. The result for $\Zcont$ is:

\begin{eqnarray}\label{eq:Zcont}
\Zcont& =&
\left(\alphacont(\mbstar)\over\alphacont(\qstar)\right)^{\gamma_0/2b_0}
\times\\
  &&\left( 1 + {\alphacont(\mbstar)-\alphacont(\qstar)\over4\pi}
  \left[{\gamma_0\over2b_0}\left\{{\gamma_1\over\gamma_0} -
  {b_1\over b_0}\right\}\right] +
{c_1\alphacont(\mbstar)\over 4\pi}\right).\nonumber
\end{eqnarray}
In this equation $\gamma_0$ and $\gamma_1$ come from the anomalous
dimension $\gamma$ of the heavy-light axial current in the effective
theory,
\[
\gamma = \gamma_0 {\alphacont\over4\pi} + \gamma_1 \left(
 {\alphacont\over4\pi} \right)^2,
\]
while $c_1$ comes from the effective to full theory matching at one
loop in the continuum, obtained by the method described at the
beginning of section~\ref{sec2intro}. This matching produces a
contribution to $\Zcont$ of
\[
1 + c_1 {\alphacont\over4\pi}.
\]
Note that Ji and Musolf~\cite{JandM} quote $c_1$ with a sign error in
the term which differs for vector and axial vector currents. The
values of $\gamma_i$ and $c_1$ are:
\[
\begin{array}{rcl}
\gamma_0 &=& -4\\
\gamma_1 &=& -254/9 - 56\pi^2/27 + 20n_f/9\\
c_1 &=& -8/3
\end{array}
\]

\begin{table}
\renewcommand{\baselinestretch}{1.0}
\centering
\caption{Length scales and renormalization constants used in this paper. The
values of $a^{-1}$ for $\b=5.7, 5.9,$ and $6.1$ are taken from
Ref. [43]. For $\beta=6.3, a^{-1}$ is estimated from that at 6.1 by
one-loop asymptotic freedom. Plaquette expectation
values are taken from [20].
$\tilde{Z}$ is the lattice to continuum renormalization factor
for the axial current, $Z_{cont}$ is the factor relating the continuum
heavy quark effective theory to full QCD, and $Z_A$ is the overall
renormalization factor used in previous discussions.}
\vspace{.1in}
\label{tab:scales}
\begin{tabular}{|c|c|c|c|c|c|}
\hline
\multicolumn{1}{|c|}{$\beta$}
&\multicolumn{1}{c|}{ $a^{-1} (\gev)$}
&\multicolumn{1}{c|}{ $\vev{\third {\rm Tr}\,{U_{\sl plaq}}}$ }
&\multicolumn{1}{c|}{ $\tilde Z$}
&\multicolumn{1}{c|}{ $\Zcont$}
&\multicolumn{1}{c|}{ $Z_A$} \\[2pt]
\hline
5.7  & 1.15(8)  & 0.549 & 0.73 & 1.00 & .63 \\
5.9  & 1.78(9)  & 0.582 & 0.77 & 0.96 & .65 \\
6.1  & 2.43(15) & 0.605 & 0.80 & 0.94 & .68 \\
6.3  & 3.08(18) & 0.623 & 0.81 & 0.93 & .68 \\
\hline
\end{tabular}
\end{table}
\vspace{0.2in}
\renewcommand{\baselinestretch}{1.5}

Using equation~(\ref{eq:Zcont}) we determine the $\Zcont$ values given
in Table~\ref{tab:scales}. These are then combined with the $\tilde
Z$'s obtained in the previous subsection to give the overall
perturbative corrections $Z$ listed in Table \ref{tab:scales}.
Here, to conform with our previous notation \cite{lat93_fb} we also list the
quantity
\begin{equation}
Z_A = \frac{\tilde{Z} Z_{cont}}{\sqrt{8\kappa_c}}
\end{equation}

\subsection{Renormalization of the Heavy Quark Mass}%
Before leaving the subject of short-distance perturbative correction
we will compute the tadpole-improved estimate of heavy quark mass
renormalization.

In the continuum
static limit neither the heavy meson mass ($M_B$) nor the renormalized
heavy quark mass ($m_b$) appears explicitly,
only the residual mass combination
$\bar\Lambda = M_B - m_b$ remains.
The precise definition of the renormalized mass parameter in the
dimensionally regularized heavy quark effective theory has been discussed
by Falk, Neubert, and Luke~\cite{FNL}.
Since the heavy quark mass renormalization is a linear divergence it
vanishes in dimensional regularization with minimal subtraction.

On the lattice, the heavy quark mass is renormalized in
the static limit.  This mass shift, $\delta m$,
is proportional to 1/a with a perturbatively calculable coefficient.
Hence the mass, $M_{eff}$, calculated for the
ground state B meson in this lattice theory
can be expressed as
\begin{equation}
\label{eq:meff}
  M_{eff} = \bar\Lambda - \delta m \label{eq:dm};
\end{equation}
in terms
the continuum residual mass $\bar\Lambda$ and
the mass shift $\delta m$.  It is clear from Eq.(\ref{eq:dm}) that the
mass $M_{eff}$ is linearly divergent
as the lattice spacing $a \rightarrow 0$.
However, it would appear that by measuring $M_{eff}$ and
removing  the tadpole-improved one-loop mass counterterm, we
have a determination of $\bar\Lambda$.  Hence we could obtain $m_b$ as
defined in Ref.~\cite{FNL} from $m_B$.
Unfortunately, Bigi {\it et al}~\cite{renormalons}
argue that non-perturbative effects ruin the preceding connection.
Even if that is so, it is still possible to verify that the linearly
divergent piece of $M_{eff}$ is correctly determined
by the perturbative calculation of $\delta m$.

At one loop, the lattice heavy quark mass renormalization
is of the form,
$$
\delta m = - {\alpha_S \over 3 \pi} {1\over a}\int {d^4 q\over \pi^2} k(q)\ .
$$
The one-loop correction to the self-energy is linearly divergent and positive.
The coefficient, $\delta m$, of the mass counterterm is negative.
The integrand k(q) is given by~\cite{EichtenHillII},
$$
k(q) = \frac{1}{8}\left[\sum_{\alpha=1}^3 \sin^2(q_{\alpha}/2)\right]^{-1} .
$$

As already noted the effect of tadpole improvement on the fitted
value of the heavy quark mass is to change
$$
C\rightarrow C+{\ln u_0\over a},
$$
This non-perturbative reduction of the mass is accompanied by a reduction
in the counterterm:
$$
\widetilde{\delta m} =  - {\alpha_S \over 3 \pi}
                       {1\over a}\int {d^4 q\over \pi^2} \tilde k(q)\ .
$$
where
$$
\tilde k(q) = k(q) + j(q)
$$
As in the calculation of the scale $q^*$, we need to compute the
expectation vlaue of $\ln q^2$ in the integral of $\tilde k(q).$
The formulae determining this scale are:
\begin{eqnarray}
X  &\equiv & \int d^4 q ~\tilde k(q) = 10.07 \\
\vev{\ln(qa)^2} &\equiv & {1\over X} \int d^4 q ~\tilde k(q) \ln(q^2a^2)
= {14.34\over10.07}\\
p^*a &\equiv & \exp[\vev{\ln(qa)^2}/2] = 2.04\ .
\end{eqnarray}
The calculational technique used is exactly that of Ref.~\cite{HernHill},
and the errors on the numerically evaluated
values of $X$ and $X \ln<q^2>$
are order~1 in the last decimal place.

The values of $\alpha_V(p^*)$ at the $\beta$ values used in this
study and $a\widetilde{\delta m} + \ln u_0$ are presented in
Table~\ref{tab:deltamcorrection}.

\begin{table}[htb]
\renewcommand{\baselinestretch}{1.0}
\centering
\caption{Tadpole improved mass counterterm for a static
quark at the $\beta$ values used in this paper. The associated gauge
couplings $\alpha_V(p^*)$ are also listed. The scale $p^*$ is $2.04/a$
(see text).}
\vspace{.1in}
\label{tab:deltamcorrection}
\begin{tabular}{|c|c|c|c|c|}
\hline
\multicolumn{1}{|c|}{$\beta$}
&\multicolumn{1}{c|}{$\alpha_V(p^*)$}
&\multicolumn{1}{c|}{$a\widetilde{\delta m} + \ln u_0$} \\[2pt]
\hline
5.7  & 0.228 & -0.394 \\
5.9  & 0.191 & -0.339 \\
6.1  & 0.170 & -0.307 \\
6.3  & 0.156 & -0.285 \\
\hline
\end{tabular}
\end{table}
\renewcommand{\baselinestretch}{1.5}

Two methods for the determination of $\bar\Lambda$ are possible.
The first of these is to do the tadpole-improved perturbative
subtraction just
described.  This results in a statistically independent result
for $\bar\Lambda$ for each $\beta$ value.  The second method
is to use the fact that the subtracted term is proportional to $1/a$, and
the physical value desired is independent of the lattice spacing.  A
two-parameter fit of the mass over the four values of $\beta$ has two
degrees of freedom, and one of the parameters is the one we desire.
The latter procedure ignores the running of the coefficient
of the 1/a term.

\subsection{Comparison and Summary}%
It is worthwhile at this point
to compare the results in Table~\ref{tab:scales}
with the widely used value of~$Z_A$ of~$0.8$,
which does not benefit from tadpole improvement.
Consider the results at $\beta=6.1$.
{}From the table, we find $\tilde Z \Zcont / \sqrt{8\kappac}$ to be 0.68.
Consequently, tadpole-improved perturbation theory results in a
reduction of the physical value of $f_B$ by a factor of $0.68/0.8$,
{\it i.e.}, a reduction of~18\%.

As an aside, we note that for the clover action---also termed ``improved'',
but in the sense that order $a$ effects rather than tadpole graphs are
being incorporated---the axial current has been
renormalized~\cite{BorrelliAndPittori}, but tadpole improvement has not
been applied to this operator.

Now that perturbation theory has been reorganized to
include tadpole corrections to all orders, we expect our one-loop calculation
of the renormalization factor to be accurate to
about 7\%.
This estimate of the magnitude of the two-loop corrections is
obtained simply by squaring
the largest one-loop correction for the various values of
$r$ quoted in Ref.~\cite{HernHill}, for both discretizations of the
axial current considered there.

While the values for $\Zcont$ depend on the estimate of the lattice spacing,
this dependence is weak; an increase by~10\% in the estimate of the scale
results in a reduction of at most~1.5\% in the value of $\Zcont$.  Thus this
source of uncertainty in the $Z$ factor is negligible compared to both the
estimated size of the two-loop corrections or the direct dependence of
$f_B$ on the lattice spacing.


\newpage
\section{Analysis Procedure for Multistate Smearing}

\subsection{Relativistic Quark Model for Heavy-light Systems}
   The rapid deterioration\cite{noise,lat91_multistate}
of the signal to noise quality of Euclidean
correlators of heavy-light mesons at large Euclidean time makes the
choice of an efficient smearing scheme essential if we wish to
extract accurately the properties of low-lying heavy-light systems.
In the multistate smearing approach previously
introduced\cite{lat91_multistate}, the
coupling of smeared bilocal Coulomb gauge operators to higher meson
states was reduced by using smearing wavefunctions obtained from
a relativistic quark model (RQM). The basic features of such a model
are (a detailed examination of the connection of such a model
with the full field theory in the case of the t'Hooft model can be
found in \cite{Schnapka};see also,\cite{cea}): \\
(a) the use of a relativistic kinetic term $\sqrt{p^{2}+\mu^{2}}$ (with
$\mu$ a constituent quark mass) for the kinetic piece of the
Hamiltonian, and \\
(b) a static confining potential $V(\vec{r})$, which can be chosen to
be the static interaction energy obtained from correlators of
temporal Wilson lines in lattice QCD.

  The importance of relativistic kinematics in determining the shape
of meson wavefunctions (with a light quark) was already implicit, if
not clearly recognized, in the puzzling persistence of purely
exponential falloff ($\simeq \exp{-Cr}$) of hadronic wavefunctions,
instead of the more rapid falloff one might naively expect
in a confining model ($\simeq \exp{-Cr^{3/2}}$ for a nonrelativistic
particle in a linearly
rising potential). This exponential falloff is due to the nonlocal character
of the kinetic part of the RQM Hamiltonian
\begin{eqnarray}
  K(\mid r-r^{\prime}\mid)&\equiv&\int\sqrt{p^{2}+\mu^{2}}\exp{i\vec{p}
  \cdot(\vec{r}-\vec{r}^{\prime})}d^{3}p \\
  &\simeq& \mid r-r^{\prime}\mid^{-9/2}\;e^{-\mu\mid r-r^{\prime}\mid},
  \;\;\;\;\mid r-r^{\prime}\mid
  \rightarrow \infty
\end{eqnarray}
which implies that $\Psi_{n}(r)$ satisfying
\begin{equation}
\label{eq:relpsi}
\int d^{3}r^{\prime}K(\mid r-r^{\prime}\mid)\Psi_{n}(r^{\prime})
+V(r)\Psi_{n}(r)=E_{n}\Psi_{n}(r)
\end{equation}
{\em cannot fall exponentially faster} than $e^{-\mu r}$ (if it did, the
integral over
$r^{\prime}$ in (\ref{eq:relpsi}) would be dominated by $r^{\prime}\simeq 0$,
giving
an asymptotic behavior $\simeq e^{-\mu r}$ for the kinetic term, in
contradiction with the assumed asymptotic behavior of $(E_{n}-V(r))
\Psi(r)$ for {\em any} $V(r)$ with power growth). In other words,
irrespective of the power rise of the confining potential,
relativistic kinematics automatically smears out the wavefunction of
a light quark over the Compton wavelength corresponding to the
constituent quark mass.

  In the static limit in which the heavy quark mass is taken to infinity,
the relativistic Schrodinger equation (\ref{eq:relpsi}) gives a single
parameter
fit (the constituent quark mass $\mu$ is the only adjustable parameter after
the static potential has been measured on the lattice) to a complete set
of orthogonal spin-independent wavefunctions corresponding to arbitrary
radial and orbital excitations of the heavy-light system. To minimize
lattice discretization and finite volume artifacts in the comparison
of RQM and lattice Monte Carlo results, we have generated a set of
lattice smearing functions by solving a discretized version of
(\ref{eq:relpsi}),
in each case on lattices of the same size as those used in the Monte
Carlos, and in each case with the static potential determined from
Wilson line correlators in the same gauge configurations used to extract
our quenched QCD results. Namely, on each lattice and for each $\beta$
value, we have extracted a full lattice static potential $V(\vec{r})$
by measuring the correlator of two Wilson lines of time extent $T$
(in Coulomb gauge-fixed configurations), and separated by a spatial distance
$\vec{r}$. The potential is then extracted by going
out in Euclidean time $T$ until the static energy
${\cal E} = -\frac{1}{T}\ln<W(0,T)W^{\dagger}(
\vec{r},T)>\equiv V(\vec{r})$ stabilizes (for example, with $\beta=$5.9
on a 16$^{3}$ lattice, this occurs for $T\geq 5$). The static potential
extracted at various $\beta$ values and lattice sizes is displayed in
Fig[\ref{fig:pot_all}].

\begin{figure}[p]
\label{fig:pot_all}
\end{figure}

  The procedure used for generating lattice smearing functions from the
RQM is as follows. We wish to obtain orthonormal lattice wavefunctions
which are eigenstates of a lattice RQM Hamiltonian defined on a
$L^{3}$ lattice (with $\vec{r},\vec{r}^{\prime}$ lattice sites):

\begin{eqnarray}
  H_{\vec{r}\vec{r}^{\prime}}&\equiv& K_{\vec{r}\vec{r}^{\prime}}
  +V(\vec{r})\delta_{\vec{r}\vec{r}^{\prime}}   \\
  K_{\vec{r}\vec{r}^{\prime}}&=&\frac{1}{L^{3}}\sum_{\vec{p}}
  \sqrt{4\sum_{i}\sin^{2}(\frac{\pi p_{i}}{L})+\mu^{2}}\;\;e^{i\vec{p}
  \cdot(\vec{r}-\vec{r}^{\prime})}
\end{eqnarray}
Such an eigenstate, in a channel of given orbital quantum number
(S,P,D etc), will correspond to a pole of the resolvent applied to
a source wavefunction $\Psi^{(0)}$ of the same orbital symmetry:
\begin{equation}
 R\equiv \|\sum_{r^{\prime}}(\frac{1}{E-H})_{\vec{r}\vec{r}^{\prime}}
 \Psi^{(0)}(\vec{r}^{\prime})\|\rightarrow\infty
 \end{equation}

 For the starting source functions $\Psi^{(0)}$, one may take for example
 a monopole localized at the origin for S-states, a dipole for P-states, etc.
 After the energy $E$ is tuned close to an eigenvalue $E_{n}$ (until
 $R$ is at least 10$^{3}$ larger than the background value), a smearing
 eigenstate $\Psi^{(a)}_{\rm smear}(\vec{r})$ is extracted by
 renormalizing the vector $\frac{1}{E-H}\Psi^{(0)}$ to unit norm. The
 inversion of $E-H$ is performed by the conjugate gradient algorithm,
 with the multiplication of the kinetic term done in momentum space
 using a fast Fourier transform.

   In most cases we have found it adequate to fit the RQM constituent mass
by matching the 1S wavefunction generated by the above procedure to the
Coulomb gauge Bethe-Salpeter wavefunction obtained at a roughly fixed
Euclidean time (corresponding to time slice 4 at $\beta$=5.9). For the
particular case of $\beta$=5.9, $\kappa$=0.159, on a 16$^{3}$ lattice,
a more detailed fitting procedure was used to determine the optimal
choice of quark mass $\mu$ in order to fit the meson wavefunction at
various time slices. The mean square deviation of $\Psi^{(1S)}_{\rm
smear}$ from the measured quenched wavefunctions at various times $T$,
for various $\mu$, is displayed in Table~\ref{tab:smearfit}.

\begin{table}[htb]
\centering
\renewcommand{\baselinestretch}{1.0}
\caption{Mean square deviation of RQM smearing and LQCD heavy-light
wavefunctions (x$10^{6}$), for $\beta=5.9, \kappa=.159$ as a function
of time. $\mu$ is the constituent quark mass parameter in the RQM
Hamiltonian.}
\vspace{.1in}
\label{tab:smearfit}
\begin{tabular}{|c|c|c|c|c|c|}
\hline
\multicolumn{1}{|c|}{$\mu$}
&\multicolumn{1}{c|}{ $T=4$}
&\multicolumn{1}{c|}{ $T=5$}
&\multicolumn{1}{c|}{ $T=6$}
&\multicolumn{1}{c|}{ $T=7$}
&\multicolumn{1}{c|}{ $T=8$} \\ \hline
0.05  & 3.37  & 1.99  & 1.36  & 5.65 & 5.95 \\
0.10  & 0.75  & 0.42  & 0.20  & 1.86 & 2.14 \\
0.12  & 0.37  & 0.45  & 0.36  & 1.02 & 1.30 \\
0.15  & 0.40  & 1.06  & 1.15  & 0.39 & 0.67 \\
0.20  & 1.75  & 3.30  & 3.61  & 0.72 & 1.00 \\
\hline
\end{tabular}
\end{table}
\renewcommand{\baselinestretch}{1.5}

  From Table~\ref{tab:smearfit} we see that the optimal choice for the
constituent
quark mass varies in the range 0.10-0.15 if we fit to meson
wavefunctions on time slices 4 to 8. We have chosen $\mu$=0.12 as the best
compromise for $\beta$=5.9, $\kappa$=0.159. With this parameter fixed,
we have generated, by the procedure outlined following (\ref{eq:relpsi}), 1S,
2S, 3S
and 4S smearing functions to be used in the multistate analysis described
in the next section. For other $\beta, \kappa$ values, we have
usually used two smearing states
only. The careful tuning of the quark mass performed here reduces to
a very small level (4 percent or less) the coupling of the exact ground
state to the higher smearing states, but will not turn out to be essential
to the extraction of accurate masses and couplings for the ground state.
A detailed discussion of
the dependence of the results on the RQM mass parameter chosen for smearing
is given in Section 5.1.

  As described in a recent article \cite{DET_RQM}, the RQM gives a
single-parameter
fit to all the excited radial and orbital meson wavefunctions of our
heavy-light system. After fixing $\mu$ by a match to the 1S wavefunction,
we have found remarkable agreement with measured excited state wavefunctions
(for example, the 1P state, cf \cite{DET_RQM}). This agreement suggests that
this
Ansatz accurately describes at least the valence quark sector of the full
mesonic bound-state.

\subsection{Multistate Smearing}

  Our object in this section is to outline a general procedure for
extracting the maximum usable information from the multistate correlator
matrix:
\begin{equation}
\label{eq:multi}
C^{ab}(T)\equiv\sum_{\vec{r}\vec{r}^{\prime}}\Psi^{(a)}_{\rm smear}(\vec{r})
<0\mid q(\vec{r},T)\bar{Q}(0,T)Q(0,0)\bar{q}(\vec{r}^{\prime},0)
\mid 0>\Psi^{(b)}_{\rm smear}(\vec{r}^{\prime})
\end{equation}
where $q (Q)$ are light (heavy) quark operators,
and the $\Psi^{(a)}_{\rm smear}$ ($a$=1,2,...$N$) contain the set of
orthonormal smearing functions obtained from the RQM as described in
the preceding section. From a set of $N_{c}$ decorrelated gauge
configurations, we begin with a corresponding ensemble of $N_{c}$
statistically independent $C^{ab}(T)$ matrices, from which a standard
deviation matrix $\sigma^{ab}(T)$ can be obtained directly. In addition
to the smearing wavefunctions of the relativistic potential model, the
set $\{\Psi^{(a)}\}$ also includes the local source generating
the desired heavy-light axial-vector matrix element for extracting $f_{B}$.
Other types of smearing (cube, wall, etc) may also be included to
facilitate an objective comparison with other recent calculations. In
(\ref{eq:multi}),
the heavy and light quark propagators in each gauge configuration are
computed in Coulomb gauge. As we are dealing with global color
singlet states on each time slice (color sums are suppressed) $C^{ab}$
is well-defined and non-zero.

  Defining states
\begin{equation}
\label{eq:smear}
 \mid \Phi^{a},T>\equiv\sum_{\vec{r}}\Psi^{(a)}_{\rm smear}(\vec{r})
 Q(0,T)\bar{q}(\vec{r},T)\mid 0>
\end{equation}
we have
\begin{eqnarray}
 C^{ab}(t)&=&<\Phi^{a},T\mid \Phi^{b},0>  \\
 &=&\sum_{n=1}^{M}e^{-E_{n}T}<\Phi^{a}\mid n><n\mid \Phi^{b}>
 +O(e^{-E_{M+1}T})
\end{eqnarray}
where the states $\mid n>$ are exact eigenstates of the lattice Coulomb
gauge transfer matrix with eigenvalues $e^{-E_{n}}$. The remainder
term of order $e^{-E_{M+1}T}$ will of course be small at large Euclidean
time, but in addition should have a small prefactor to the extent that our
smearing functions $\Psi^{(a)}_{\rm smear}(\vec{r})$ ($a$=1,2,..$M$) do a
good job in representing the valence quark structure of the low-lying
states, and to the extent that more complicated Fock states (containing
extra quark pairs, real gluons, etc) are not too important.

  Next, define mixing coefficients (in our case, they are real):
\begin{equation}
  v^{a}_{\;\;n} \equiv <\Phi^{a}\mid n> = <n\mid\Phi^{a}>
\end{equation}
Neglecting the exponential contamination of order $e^{-E_{M+1}T}$, we see
that the multistate coupling matrix can be fit to an expression of the
form
\begin{equation}
\label{eq:multistate_formula}
  C^{ab}(T)=\sum_{n=1}^{M} v^{a}_{\;\;n}v^{b}_{\;\;n}e^{-E_{n}T}
\end{equation}
Of course, we cannot hope to extract $M$ independent time-dependencies
with $N<M$ smearing wavefunctions, so only $N\geq M$ will be considered.
Typically we shall extract the maximum information from the lattice data
by picking $N=M+1$ (the extra source function being the local current needed
for the extraction of $f_{B}$).

  The fit is performed by a chi-square minimization of
\begin{equation}
\label{eq:chisquared}
  \chi^2\equiv\sum_{a,b}\sum_{T=T_{<}}^{T_{>}}
  \frac{\mid C^{ab}(T)-\sum_{n=1}^{M}v^{a}_{\;\;n}v^{b}_{\;\;n}e^{-E_{n}T}
  \mid^{2}}{\sigma^{ab}(T)^{2}}
\end{equation}
with respect to the fitting parameters $\{ v^{a}_{\;\;n},E_{n} \}$, over
a fitting range $T_{<} \leq T \leq T_{>}$ in Euclidean time. The fit
is performed on an ensemble of $N_{c}$ jack-knife coupling
matrices obtained by replacing each in turn of the $N_{c}$ coupling matrices
by the average matrix and reaveraging. We have chosen $\mid \Phi^{N}>
\equiv J_{\rm axial}(0)\mid0>$, so the parameters $v^{N}_{\;\;n}$ should
be interpreted as {\em lattice} f-parameters for the ground and excited
meson states, $E_{n}$ as the corresponding masses, and $v^{a}_{\;\;n}$
($a$=1,2,..,$M$)
as mixing coefficients indicating the degree of overlap of the exact meson
states with our RQM smeared states $\mid \Phi^{a}>$.
The sum over $a,b$ in (\ref{eq:chisquared})
does {\em not} include the local-local correlator $a=b=N$, which is not
well described by a sum over a few low-lying states. Note that this fitting
procedure automatically gives the lattice f-parameters without the need
to divide by the square-root of the smeared-smeared correlator as in the
usual approach. Moreover, the ensemble of $N_{c}$ parameter sets
$\{ v^{a}_{\;\;n},E_{n} \}$ obtained in this way can be subjected to a
straightforward statistical analysis to determine the error in each of
these parameters separately, correlations between parameters (e.g. between
masses and f-parameters), and so on.

  Once the overlaps $<\Phi^{a}\mid n>$ have been estimated by a best fit
of $C^{ab}(T)$, a smearing operator can be constructed which is guaranteed
to contain {\em at most one} of the first $N$ exact meson states, thereby
removing any other exponential time-dependence to the $e^{-E_{N+1}T}$
level. Specifically, if $\epsilon_{a_{1}a_{2}..a_{N}}$ is the totally
antisymmetric symbol in $N$ dimensions, the smeared state
\begin{equation}
\label{eq:antisymm}
\mid \hat{\Phi}^{A}>\equiv\epsilon_{a_{1}a_{2}..a_{N}}\prod_{i\neq A}
v^{a_{i}}_{\;\;\;i}\mid\Phi^{a_{A}}>,\;\;\;\;A=1,2,..M
\end{equation}
is guaranteed (to the extent that we have accurately extracted the
mixing coefficients $v^{a}_{\;\;n}$) to contain only the exact meson
state $\mid A>$, together with contaminations from the $(N+1)$'th excited
state and higher. An effective mass plot of the usual kind can then
be obtained for the A'th state by displaying (we use a smeared-local
correlator to minimize noise)

\begin{equation}
  m^{A}_{\rm eff}(T)\equiv\ln{\frac{C^{A}(T-1)}{C^{A}(T)}}
\end{equation}
where
\begin{equation}
   C^{A}(T)=<\hat{\Phi}^{A},T\mid \Psi^{(\rm loc)},0>
\end{equation}
Of course, this plot will be most flat for the ground state $A=1$,
where the relative exponential contamination is reduced to the level
$e^{-(E_{N+1}-E_{1})T}$, and where small admixtures of lower-lying states
cannot creep in to distort the effective mass plateau. The
effective mass plots for the ground state at $\beta  = 5.7,
5.9, 6.1$, and $6.3$ for various kappa values are shown in
Figs.[\ref{fig:effb}-\ref{fig:effg}]. The solid line in each of these plots
represents the ground state energy extracted from the full multistate fit
over the time window indicated by the length of the line. For each $\beta$
the time window for the multistate fit was chosen to be over approximately
the same interval in physical units, viz.
about $\frac{1}{3} fm $ to $1 fm$.
Noting that the splitting between the ground state and the second
excited state is found, in our multistate calculations,
to be around $800$ to $900 MeV$, the choice of $\frac{1}{3} fm$ for the
lower end of the time window should provide an exponential supression of
excited states by at least a factor of 3. In addition, our careful tuning of
the smearing functions should produce a relatively small coefficient for
the higher excited states. The equality of the smeared-smeared and
smeared-local
effective masses exhibited in the plots confirms our choice of fitting
interval.
A more complete discussion of systematic errors due to excited states is
given in Section 5.1.

\begin{figure}[p]
\label{fig:effb}
\end{figure}

\begin{figure}[p]
\label{fig:effe}
\end{figure}

\begin{figure}[p]
\label{fig:effc}
\end{figure}

\begin{figure}[p]
\label{fig:efff}
\end{figure}

\begin{figure}[p]
\label{fig:effd}
\end{figure}

\begin{figure}[p]
\label{fig:effg}
\end{figure}

  The fitting formula (\ref{eq:chisquared}) is easily generalized
to allow a global
fit to the data at various $\kappa$ values (for fixed $\beta$): this
is essential in order to take into account correlations between the
coupling matrices at different $\kappa$ values, which would affect
our estimate for the error of the results when linearly extrapolated
to $\kappa_{c}$. Namely, (\ref{eq:chisquared}) is replaced by
\begin{equation}
\label{eq:chisquaredkappa}
  \chi^2\equiv\sum_{a,b}\sum_{\kappa}\sum_{T=T_{<}}^{T_{>}}
  \frac{\mid C^{ab}(\kappa,T)-\sum_{n=1}^{M}v^{a}_{\;\;n}(\kappa)
v^{b}_{\;\;n}(\kappa)e^{-E_{n}(\kappa)T}
  \mid^{2}}{\sigma^{ab}(\kappa,T)^{2}}
\end{equation}
where
\begin{eqnarray*}
  v^{a=N}_{\;\;n}(\kappa)&\equiv&v^{a}_{0\;\;n}+v^{a}_{1\;\;n}(\kappa^{-1}
-\kappa_{c}^{-1})  \\
  E_{n}(\kappa)&\equiv&E_{0\;n}+E_{1\;n}(\kappa^{-1}-\kappa_{c}^{-1})
\end{eqnarray*}
and the mixing coefficients $v^{a=1,..M}_{\;\;n}$ are varied freely. Note that
only the immediately physical mass and lattice-f parameters are assumed to have
the
chiral dependence on $\kappa^{-1}$: other mixing coefficients involve the model
dependent choice of smearing functions from the RQM.
The chi-square minimization  allows  the direct extraction of masses and
couplings
extrapolated to $\kappa_{c}$, as well as the slopes in $\kappa^{-1}$
of these quantities (all of which are free variational parameters in this new
global fitting procedure). The usual jackknife procedure can then be applied
to yield the correct errors on the extrapolated quantities.


\newpage
\section{Lattice Results for Heavy-Light Mesons}

\indent To extract results for masses and decay constants we have used the
set of gauge configurations and light quark propagators enumerated in
Table \ref{tab:gf_and_qp}. The light quark action we use is not $O(a)$
improved.
The four columns in Table \ref{tab:gf_and_qp} are the gauge coupling,
$\beta$; lattice size; number of gauge configurations (separated by
1000, 2000, 4000, and 4000 sweeps for $\beta$ = 5.7, 5.9, 6.1, and 6.3
respectively);
and the light quark $\kappa$ values calculated for each
configuration.

\begin{table}[htb]
\centering
\renewcommand{\baselinestretch}{1.0}
\caption{Summary of gauge configurations and light quark parameters used in
this paper. Listed are a letter used to identify each Monte Carlo run, the
$\beta$ value, lattice size, number of gauge configurations in each ensemble,
and the values of light quark hopping parameter $\kappa$ analyzed.}
\vspace{.1in}
\label{tab:gf_and_qp}
\begin{tabular}{||ccccc||} \hline
${\rm run}$&$\beta$ & ${\rm lattice}$ & ${\rm confs}$ & $\kappa $ \\
\hline
b & 5.7 & $12^3\times 24$ & 100 & $.168,.1667, .165, .161$  \\
e & 5.9 & $12^3\times 24$ & 100 & $.159, .158, .157, .154$  \\
c & 5.9 & $16^3\times 32$ & 100 & $.159, .158, .157, .156, .154$  \\
f & 5.9 & $20^3\times 40$ & 100 & $.159, .158, .157, .154$  \\
d & 6.1 & $24^3\times 48$ & 50 & $.1545, .154, .153, .151$  \\
g & 6.3 & $32^3\times 48$ & 50 & $.1515,.1513, .1510, .1500$  \\ \hline
\end{tabular}
\end{table}
\renewcommand{\baselinestretch}{1.5}
The multistate smearing analysis outlined in Section 3 provides a
powerful
method for extracting heavy-light meson parameters. Unlike single or double
exponential fits to single-channel ``smeared-smeared'' and ``smeared-local''
correlators, the fitting of the $N\times N$ matrix of correlators to
an expression of the form (\ref{eq:multistate_formula})
is highly constrained. As we will show, this
method allows a determination of $f_B$ and other heavy-light
parameters which is less prone to systematic
errors than previously applied methods. In this
section, we present our results. In Section 6.2 we compare these results
with those recently reported in Refs.\cite{BLS,Wuppertal,UKQCD,Rome}.
Most of the results presented here
were obtained from the fitting procedure discussed in Section 4 using $N=3$
and $M=2$, i.e. a $3\times 3$ matrix of correlators (2 smearing functions and
the $\delta$-function source) fit to the sum of two exponentials (always
excluding the local-local correlator from the fit). To estimate
the systematic errors associated with the fitting procedure, we have tried
varying both the shape of the smearing functions (by adjusting the RQM quark
mass parameter) and the number $M$ of smearing functions included.
These results are discussed below.

First consider the mass eigenvalue $E_1$ in Eq. (\ref{eq:chisquared}) which
describes the
leading exponential falloff of the heavy-light correlators
\begin{equation}
C^{ab}(t) \simeq v^a_1 v^b_1 e^{-E_1t}
\end{equation}
In the multistate fitting procedure, $E_1$ is the energy associated with
the ground state contribution to the correlator.
This parameter represents a combination of the
binding energy of the B-meson ground state plus a divergent
mass shift of the heavy quark. Recall that only the bare mass of the heavy
quark is removed in reducing to the effective static theory. The mass
shift induced by QCD is therefore measureable on the lattice as a non-scaling
piece in the parameter $E_1$. We measure the dependence of $E_1$ on both
the light-quark hopping parameter $\kappa$ and the lattice spacing $a$.
The graph in Fig.~\ref{fig:mb_vs_k} shows the $\kappa$-dependence
of $E_1$ for the four
values of $\beta$ studied. In each of the four data sets, the dependence on
$\kappa^{-1}$ is quite linear, allowing an accurate extrapolation to the chiral
limit $\kappa_c$. Numerical results are tabulated in
Table ~\ref{tab:hl_beta}. For each value of beta and kappa the results in this
table were obtained from a 2-state fit, with the $\chi^2$ per degree of freedom
of each fit listed in the last column. For $\beta=5.9$, the $\chi^2/dof$
are those obtained on the $20^3$ box. The results for
$\tilde{f}_B$ and $aE_1$ at $\beta=5.9$ listed in Table ~\ref{tab:hl_beta}
are the infinite volume values obtained by fitting all three box sizes to a
Luscher finite volume formula, as described in Section 5.2 (except for
$\kappa=.156$, which has only been done on the $16^3$ box). The results for the
three box sizes are listed separately in Tables ~\ref{tab:mb_vol} and
{}~\ref{tab:fb_vol}. A measure of the overlap between the true ground state and
the RQM wavefunction smeared operators is given by
$\sqrt{2\kappa}\left[\sum_{a=1}^{M}(v_1^a)^2\right]^{\frac{1}{2}}/\sqrt{6}$.
The measured values of this overlap for each of the fits is recorded in
Table ~\ref{tab:hl_beta}.

\begin{figure}[htp]
\label{fig:mb_vs_k}
\end{figure}

\begin{table}[htp]
\centering
\renewcommand{\baselinestretch}{1.0}
\caption{Lattice results for heavy-light mesons (static approximation).
Values for the ground state energy $aE_1$ and decay constant $\tilde{f}_B$
are extracted from a 2-state fit over time window $\Delta T$. ($\tilde{f_B}$
is related to the physical decay constant $f_B$ by Eq. (30)). Results
at $\kappa = \kappa_c$ are from the multi-$\kappa$ fits as discussed in
Sec. 3.2. The column labeled $overlap$ is a measure of the total overlap
between
the true ground state and the RQM wavefunction smeared operators used.}
\vspace{.1in}
\label{tab:hl_beta}
\begin{tabular}{|c|c|c|c|c|c|c|}
\hline
\multicolumn{1}{|c|}{$\beta$}
&\multicolumn{1}{c|}{$\Delta T$}
&\multicolumn{1}{c|}{ $\kappa$}
&\multicolumn{1}{c|}{ $aE_1$}
&\multicolumn{1}{c|}{ $overlap$}
&\multicolumn{1}{c|}{ $\tilde{f}_B$}
&\multicolumn{1}{c|}{ $\chi^2/dof$} \\ \hline
5.7  & 2-8 & .161  & .827(6)  & .725(7) & .670(19)& .50\\
     & & .165  & .794(8)  & .717(8) & .626(23)& .78\\
     & & .1667 & .776(9)  & .699(7) & .590(24)& .54\\
     & &.168  & .767(11) & .694(9) & .578(29)& .43\\
     & &$\kappa_c$ & .758(10) & .694(8) & .564(28)& .58\\
5.9  &3-10& .154  & .719(5)  & .742(12) & .347(11)& .87\\
     && .156  & .692(8) & .731(13) & .318(13)& 1.06\\
     && .157  & .678(6) & .724(13) & .300(11)& .62\\
     && .158  & .665(7) & .716(15) & .283(12)& .56\\
     && .159  & .645(9) & .704(18) & .259(14)& .60\\
     && $\kappa_c$ & .638(9) & .686(24) & .250(14)& .66\\
6.1  &4-12& .151  & .620(7)  & .769(17) & .199(10)& .57\\
     && .153  & .583(9) & .721(24) & .170(12)& .60\\
     && .154  & .561(11) & .705(24) & .149(12)& .61\\
     && .1545 & .551(13) & .700(30) & .142(14)& .48\\
     && $\kappa_c$ & .544(12) & .689(25) & .135(13)& .55\\
6.3  &5-14& .1500 & .528(7)  & .748(18) & .120(7)& .72\\
     && .1510 & .511(7)  & .728(20) & .107(7)& .66\\
     && .1513 & .506(8)  & .724(20) & .104(7)& .61\\
     && .1515 & .504(8)  & .729(20) & .103(8)& .56\\
     && $\kappa_c$ & .499(9) & .720(17) & .099(8)& .62\\
\hline
\end{tabular}
\end{table}
\renewcommand{\baselinestretch}{1.5}

In addition to analyzing the data at each value of $\kappa$
separately, we have also performed a simultaneous fit to all $\kappa$ values
for a given $\beta$ by allowing the multistate fitting parameters $v^a_n$ to
depend  on $\kappa$, as described in Section 3. Using this procedure on
jackknifed subensembles provides
a better estimate of the error on the chiral extrapolation, since it takes
account of the fact that the different $\kappa$ values have correlated errors.
Comparing the results of this analysis with results of a separate analysis of
each
$\kappa$, we find that the extrapolated values for $E_1$ and $f_B$ change
very little, while the error bars on these results are about 30\% lower than
those obtained by ignoring $\kappa$ correlations. On the other hand, the
results for the slope of these quantities as a function of $\kappa^{-1}$ are
greatly
improved by the simultaneous-$\kappa$ fit, reducing the errors by a factor of
3 or more over the independent-$\kappa$ analysis.  Thus it is especially
important to take account of inter-$\kappa$ correlations for quantities such
as $m_{B_s}-m_{B_u}$ and $f_{B_s}/f_{B_u}$.

Let us consider $E_1$ as a function of the lattice spacing
$a$ and of the naive light quark mass (see Appendix).
The linear dependence on $\kappa^{-1}$ becomes
\begin{eqnarray}
E_1(2am_q,a)&=& E_1(0,a) + (\kappa^{-1}-\kappa_c^{-1})E_1^{\prime}(0,a)\\
          &=& E_1(0,a) + 2am_q E_1^{\prime}(0,a) \nonumber
\end{eqnarray}
The quantity $M_{eff}$ defined in Section 2, Eq. (\ref{eq:meff}) is just
$E_1$ evaluated at $\kappa = \kappa_c$,
\begin{equation}
M_{eff} = E_1(0,a)
\end{equation}
The slope parameters $E_1^{\prime}(0,a)$ are obtained by a correlated
fit to all $\kappa$-values, as described in Section 3. The results for
the slopes and intercepts are given in Table~\ref{tab:heavy_light_crit}.

\begin{table}[htp]
\centering
\renewcommand{\baselinestretch}{1.0}
\caption{Slopes and intercepts for $aE_1$ and $\tilde{f}_B$ as a
function of $\kappa^{-1}-\kappa_c^{-1}$. For a given $\beta$,values are
obtained from a simultaneous two-state fit to all values of $\kappa$
over a time window $\Delta T$.}
\vspace{.1in}
\label{tab:heavy_light_crit}
\begin{tabular}{|c|c|c|c|c|c|}
\hline
\multicolumn{1}{|c|}{$\beta$}
&\multicolumn{1}{c|}{$\Delta T$}
&\multicolumn{1}{c|}{$aE_1(0)$}
&\multicolumn{1}{c|}{ $E_1^{\prime}(0)$}
&\multicolumn{1}{c|}{ $\tilde{f}_B(0)$}
&\multicolumn{1}{c|}{ $\tilde{f}_B^{\prime}(0)$} \\ \hline
5.7  &2-8 & .758(11)  & .236(25)  & .564(29) & .424(72)\\
5.9  &3-10& .638(9)  & .350(25)  & .250(13) & .444(38)\\
6.1  &4-12& .544(12) & .450(47)  & .138(13) & .387(46)\\
6.3  &5-14& .499(9)  & .376(56)  & .099(7) & .290(56)\\
\hline
\end{tabular}
\end{table}
\renewcommand{\baselinestretch}{1.5}
\begin{figure}[htp]
\label{fig:fb_vs_k}
\end{figure}
In a similar way, we obtain a linear fit (see Fig. \ref{fig:fb_vs_k})
 to the $\kappa^{-1}$ dependence
of the groundstate pseudoscalar decay constant $f_B$. Define a
quantity $\tilde{f}_B$ which is just the matrix element parameter
$v_1^N$ in the multistate fit (overlap of the ground state with the
$\delta$-function source), with a normalization factor $\sqrt{2\kappa}$
for the light quark included.
In the scaling limit, the physical value of $f_B$ is related to $\tilde{f}_B$
by the following multiplicative constants,
\begin{equation}
f_B = \tilde{f}_B\times \sqrt{\frac{2}{M_B}}\times
a^{-\frac{3}{2}}\times Z_A
\end{equation}
where $Z_A$ is the renormalization factor associated with matching
the full theory with the effective static theory on the lattice, as discussed
in \cite{EichtenHillII}, \cite{Boucaud}, and in Section 2.

The lattice spacing dependence of $E_1$ in the chiral limit $m_q=0$ is
plotted in Fig. \ref{fig:mb_vs_a}. The results are consistent with a linear $a$
dependence,
\begin{equation}
aE_1(0,a) = E_1(0,0) + \tilde{E}_1 a
\end{equation}
\begin{figure}[htp]
\label{fig:mb_vs_a}
\end{figure}

with
\begin{eqnarray}
E_1(0,0) = .351(14)\\
\tilde{E}_1 = .481(25) {\rm GeV}
\end{eqnarray}
The first term $E_1(0,0)$ is the linearly divergent (i.e. $O(1/a)$) term in the
heavy quark mass shift.

The tadpole-improved estimate of $a \delta m$ was discussed in Section~2.
The corresponding quantity is $a \widetilde{\delta m} + \ln u_0$, and
was tabulated in Table \ref{tab:deltamcorrection}.
Comparing the ground state effective mass with
the  tadpole improved 1-loop result at each lattice spacing
one finds about a 30 \% discrepancy in the singular part of
the mass shift, which can easily be accounted for by higher-loop and/or
nonperturbative contributions.
In fact, ordinary (non-improved) one-loop perturbation theory
\cite{EichtenHillII} gives
\begin{equation}
\delta m = - \frac{1}{a}\times\frac{g^2}{12\pi^2}\times 19.95
\end{equation}
Simply identifying this value with the extrapolated lattice result
gives $\alpha_s \simeq 0.162(6)$ which is in reasonable agreement with
other determinations of $\alpha_s$ in the range of lattice spacings considered
here \cite{LepMack}. So the entire ``discrepancy" can be removed by a
reasonable
redefinition of the perturbative coupling being used.

The mass of the meson $B_s$, composed of a $b$ quark and a strange antiquark,
is of considerable phenomenological interest. Our calculation of the
heavy-light ground state energy as a function of $\kappa$ provides a
determination of the mass splitting between the $B_s$ and the $B_u$
mesons
\begin{equation}
\Delta M_{B_s} = M_{B_s} - M_{B_u}
\end{equation}
Since the divergent self-mass of the heavy quark is independent of light quark
mass, it will cancel in the mass difference $\Delta M_{B_s}$, and the latter
should therefore scale properly with $a$. For each value of $\beta$, we use the
determination of $\kappa_s$ and $\kappa_u$ discussed in the Appendix, along
with
the observed $\kappa$ dependence of $E_1$ to determine $\Delta M_{B_s}$. The
results are shown in Fig. \ref{fig:mbstrange_splitting_vs_a}.
\begin{figure}[htp]
\label{fig:mbstrange_splitting_vs_a}
\end{figure}

A linear extrapolation of the mass difference $\Delta M_{B_s}$ to $a=0$ gives
\begin{equation}
\Delta M_{B_s} = 86\pm 12\, \mev
\end{equation}
Notice that the results for $\Delta M_{B_s}$ shown in
Fig. \ref{fig:mbstrange_splitting_vs_a} exhibit a fairly mild dependence on
the lattice spacing, in marked contrast to the strong $a$-dependence of $f_B$.
The decay constant $f_{B_s}$ for the strange B-meson may be determined,
using the values for the slope parameter $\tilde{f}_B^{\prime}$ in Table
\ref{tab:heavy_light_crit}. The ratio $f_{B_s}/f_{B_u}$ is plotted in
Fig. \ref{fig:fbs_fbu_vs_a}. Again, the $a$-dependence of the ratio is
much weaker than that of $f_B$ itself. Extrapolating to $a=0$, we obtain
\begin{equation}
\frac{f_{B_s}}{f_{B_u}} = 1.216 \pm 0.041
\end{equation}
\begin{figure}[htp]
\label{fig:fbs_fbu_vs_a}
\end{figure}


\newpage
\section{Determination of Systematic Errors}

\subsection{Systematic error due to excited state contamination}

\indent The main results presented for $f_B$ have been obtained by a 2-state
fit
to the correlators which employed quark-model smearing functions for the
ground state (1S) and first radially excited (2S) state. As pointed out
in Section 3, this produced quite stable effective mass plots which indicated
that accurate ground state parameters could be extracted with time separations
as short as $T=2$ or 3. Since the errors in most previous calculations have
been
dominated by the systematic effect of higher state contamination, it is
particularly important to estimate the size of this effect
to get an overall determination of the accuracy of our results. To further
investigate this issue, we have carried out a more
complete study of the dependence of the extracted $\tilde{f}_B$ value on the
fitting
procedure. First, we have varied the size of the source smearing function
by changing the quark mass parameter $\mu$ in the RQM wavefunctions.
We then compare the results from the 2-state fit with those from a truncated
1-state fit (using only the correlators of the ground state smearing function
and the delta-function source).

\begin{figure}[htp]
\label{fig:mb_1s2s}
\end{figure}

In Figure \ref{fig:mb_1s2s} we compare the effective mass from
the 1-state fit with that of the 2-state fit.
After determining the 1S-2S splitting $\Delta$ from a
2-state fit, the effective masses obtained over a Euclidean time window can
be plotted versus the variable $e^{-\Delta t}$, allowing an extrapolation
to $t=\infty$ (See below).
The results of the 1-state fit are plotted for four different time windows,
1-6, 2-7, 3-8, and 4-9, and for four different choices of RQM smearing
functions (with quark mass $\mu = .32, .60, .90,$ and $1.20$). (Note: The
result from window $t_{<}-t_{>}$ is plotted at the value of
$e^{-\Delta t}$ corresponding to $t = t_{<}$.)
The effective masses from the 2-state fit using the four different
$\mu$ values and the time window 3-10 are all plotted on the far left side
of the plot at $e^{-\Delta t}=0$. The 1-state results for the different
smearing function choices are clearly converging to a common effective mass
at $t=\infty$ which agrees well with the 2-state result, the latter being
quite insensitive to the choice of $\mu$.
Similarly, the result for $f_B$ from the full 2-state
fit remains unchanged, within errors, for a wide variation of the $\mu$
parameter. On the other hand, the result from the 1-state fit varies by
20-30 \% over the same range of $\mu$ values. This provides strong evidence
that the 2-state fit does a good job of isolating the ground state, even when
the chosen smearing functions are not very well tuned. To demonstrate
this, we look at the dependence of these results on the time window chosen
for fitting. While the two-state fits are generally stable under variation
of the time window, the 1-state fit shows a systematic time-dependence.
If this time-dependence is assumed to be largely
due to contamination from the first excited
state, it should fit asymptotically to the functional form
\begin{equation}
\label{eq:1-state}
\tilde{f}_B(t) = \tilde{f}_B(\infty) + C e^{-\Delta t}
\end{equation}
where $\Delta = E_2 - E_1$ is the energy splitting between the ground state
and first excited state. Without an independent estimate of this
splitting, it is difficult to obtain a reliable determination of the parameters
in (\ref{eq:1-state}) directly from the results of the 1-state fit.
On the other hand, the 2-state fit determines both $E_1$ and $E_2$, and
therefore $\Delta$. If we use this determination to fix $\Delta$, the formula
(\ref{eq:1-state}) can be used to extrapolate the 1-state results to
$t = \infty$. Comparing this result with that of the full 2-state fit
provides a useful and nontrivial check on the assertion that the systematic
effect from excited states is under control. Figure \ref{fig:fb_1s2s} shows
the results of such a comparison for the case $\beta=5.7$,$\kappa = .161$.
The splitting
obtained from the 2-state fit is $\Delta = .321$ in lattice units. It is seen
in Fig. \ref{fig:fb_1s2s} that, for the time window 1-6 (far right on
the graph),
the result for $\tilde{f}_B$ varies systematically with the choice of smearing
function. As $t$ gets larger, the results from the different smearing functions
tend to converge to the same value. The points plotted at $e^{-\Delta t}=0$
include the four extrapolated values obtained from Eq. (\ref{eq:1-state}).
For comparison, the results of the 2-state fit for the four $\mu$ values
and $t= 3-10$ are also plotted.
All of these points are well within a standard deviation of each other.
Similar comparison of the results of 1-state and 2-state fits for other values
of $\beta$ and $\kappa$ give comparable agreement.
Based on this agreement, we conclude that the systematic error on our
results due to excited states has been eliminated at the level of our present
statistics.

\begin{figure}[htp]
\label{fig:fb_1s2s}
\end{figure}

It is worth emphasizing here that our ability to control excited state
contamination depends crucially on the use of the multi-state fitting
procedure. Although the 1-state fits were all found to lead to consistent
results after extrapolation to $t = \infty$, an accurate extrapolation
would not have been possible without an independent determination of the
splitting $\Delta$, which is only obtainable from the 2-state fit.

\subsection{Finite volume corrections}

\indent Using the scales in Table \ref{tab:scales}, we find that the
physical volumes of the boxes for the main ensembles
used in our calculation are approximately
$(2.0 fm)^3$, $(1.8 fm)^3$, $(1.9 fm)^3$, and $(2.0 fm)^3$ for $\beta =
5.7\,(12^3)$,
$5.9\,(16^3)$, $6.1\,(24^3)$, and $6.3\,(32^3)$ respectively. Although these
volumes appear
to be comfortably large compared to the observed size of the ground
state B-meson on the lattice, we consider in this section the possibility
of corrections to our results due to finite volume effects
and describe our method for estimating these effects. Although our overall
conclusion is that these effects are negligible on the lattices considered
here, the estimates discussed in this section may be
useful for selecting $\beta$'s and lattice sizes in subsequent
studies.

An extensive theoretical study of finite volume effects on field
theoretic calculations has been carried out by Luscher \cite{LuscherI}.
Consider, for example, the effect on the mass of a particle $m_P$. For
large enough volume, the leading effect is due to the propagation of the
lightest mass meson (e.g. pion) ``around the world,'' leading to the
expression for $m(L)$, the particle mass in an $L\times L \times L$ box,
\begin{equation}
\label{eq:finite_volume}
m_P(L) = m_P(\infty) + A\frac{e^{-\lambda L}}{L}
\end{equation}
where the exponent $\lambda$ is determined by the mass of the pion, and
$A$ is given in terms of the on-shell $\pi PP$ coupling. This finite
volume correction can be interpreted as the effect of squeezing the pion
cloud surrounding the particle. A somewhat different situation takes
place when the particle $P$ is a loosely bound state of constituents. In
this case, the finite size effect is caused by the squeezing of the
bound state wave function \cite{Hochberg}. As pointed out by Luscher
\cite{LuscherII}, this situation falls into the same general framework
as that which led to Eq.(\ref{eq:finite_volume}), except that, in this case,
the particle that travels around the world is one of the constituents of
the bound state. In fact, for a nonrelativistic bound state in a
non-confining potential, the finite
volume effect assumes exactly the same form as (\ref{eq:finite_volume}),
but in this case, the exponent $\lambda$ is related to the binding energy (and
hence to the spatial extent of the bound state wave function).

For the case $\beta = 5.9$ we have carried
out a complete Monte Carlo investigation of the heavy-light propagators
on lattices of three different sizes, $12^3\times 24, 16^3\times 32,$ and
$20^3\times 40$ (runs e, c, and f in Table\ref{tab:gf_and_qp}). With the value
$a^{-1}=1.78$ GeV, these three boxes are of
spatial length 1.3 fm, 1.8 fm, and 2.2 fm, respectively.
The results for
both the ground state energy $aE_1$ and for $\tilde{f}_B$ are given in
Tables \ref{tab:mb_vol} and \ref{tab:fb_vol}.
They are seen to be the same,
within errors, in all three size boxes, and thus, no significant finite
size effect is observed.
In order to determine an upper limit on the finite volume corrections to our
results, we will make the assumption that these
effects can be parametrized in the
Luscher form (\ref{eq:finite_volume}).
\begin{equation}
\label{eq:vol_E1}
aE_1(L) = aE_1(\infty) + A_{aE_1}\frac{e^{-\lambda L}}{L}
\end{equation}
\begin{equation}
\label{eq:vol_fB}
\tilde{f}_B(L) = \tilde{f}_B(\infty) + A_{\tilde{f}_B}\frac{e^{-\lambda L}}{L}
\end{equation}
(Note: For particle masses in
full QCD in a sufficiently large box, such an expression has been derived
rigorously. For masses in quenched approximation and for decay constants,
it's validity is not established, but we adopt it as a convenient ansatz.
An alternative power law form is also discussed at the end of this section.)
For a given choice of the exponential
parameter $\lambda$, a fit to Eq. (\ref{eq:finite_volume}) gives a limit
on the coefficient $A$.

\renewcommand{\baselinestretch}{1.0}

\begin{table}[htp]
\centering
\renewcommand{\baselinestretch}{1.0}
\caption{Volume dependence of $aE_1$ at $\beta = 5.9$. Results are obtained
from
a 2-state fit over time window $\Delta T = 3-10$ on lattices of size
$12^3\times 24, 16^3\times 32,$ and $20^3\times 40$. Numbers in square
brackets are the $\chi^2$ per degree of freedom for each multistate fit.}
\vspace{.1in}
\label{tab:mb_vol}
\begin{tabular}{|c|c|c|c|c|c|}
\hline
\multicolumn{1}{|c|}{$\kappa$}
&\multicolumn{1}{c|}{ $aE_1(12)$}
&\multicolumn{1}{c|}{ $aE_1(16)$}
&\multicolumn{1}{c|}{ $aE_1(20)$} \\ \hline
 .154  & .716(9)  & .718(7) & .719(6)\\
       & [.74]    & [1.12]  & [.87]  \\
 .157  & .675(12) & .679(8) & .677(8)\\
       & [.92]    & [1.01]  & [.62]  \\
 .158  & .661(14) & .667(9) & .662(9)\\
       & [.99]    & [.92]   & [.56]  \\
 .159  & .652(21) & .652(12)& .641(11)\\
       & [.98]    & [.76]   & [.60]   \\
$\kappa_c$ & .638(17) & .643(11)& .634(11)\\
           & [1.01]   & [.92]   & [.66]  \\
\hline
\end{tabular}
\end{table}
\renewcommand{\baselinestretch}{1.5}

\begin{table}[htp]
\centering
\renewcommand{\baselinestretch}{1.0}
\caption{Volume dependence of $\tilde{f}_B$ at $\beta=5.9$. (See caption of
Table 7.)}
\vspace{.1in}
\label{tab:fb_vol}
\begin{tabular}{|c|c|c|c|c|c|}
\hline
\multicolumn{1}{|c|}{$\kappa$}
&\multicolumn{1}{c|}{ $\tilde{f}_B(12)$}
&\multicolumn{1}{c|}{ $\tilde{f}_B(16)$}
&\multicolumn{1}{c|}{ $\tilde{f}_B(20)$} \\ \hline
 .154  & .341(21) & .347(13) & .346(14)\\
 .157  & .299(22) & .303(14) & .298(15)\\
 .158  & .284(24) & .288(15) & .278(16)\\
 .159  & .279(35) & .271(18) & .252(17)\\
$\kappa_c$ & .261(29) & .260(18) & .245(18)\\
\hline
\end{tabular}
\end{table}
\renewcommand{\baselinestretch}{1.5}

\renewcommand{\baselinestretch}{1.5}

Our strategy is
to extract an estimate of the exponent $\lambda$ in the Luscher formula
by two methods: (1) a direct study of the finite volume effects in the
relativistic quark model, and (2) a study of the exponential falloff of the
ground state wave function obtained in the LQCD calculation.
For both $f=f_B$ and $f=aE_1$, the results from the RQM were calculated on
$12^3, 16^3,$ and $20^3$ boxes. (The relativistic Van Royen-Weisskopf
formula\cite{cea} was used to obtain $f_B$ from the RQM wave function.)
For the lightest quark mass studied, the RQM estimate gives
$\lambda/a = 0.9$ GeV. (The results for $aE_1$ and
for $\tilde{f}_B$ are both well
fit with the same value of $\lambda$.) A slight increase in the value of
$\lambda$ for larger light quark mass is observed, but is inconsequential
for our analysis. A direct study of the exponential falloff of the LQCD
ground state wave function (using the
bound state interpretation of the
Luscher formula) gives a similar, but somewhat smaller estimate of
$\lambda/a = 0.75$ GeV for the lightest quark mass.
A smaller value of $\lambda$  assumes a slower
falloff with box size and thus allows for a larger
finite size effect on the $16^3$ and $20^3$ lattices. Thus, in order to
obtain a conservative upper bound on these effects, we have assumed a
value $\lambda/a = 0.7$ GeV, i.e. slightly smaller than the RQM and
wave function estimates.
To determine the sensitivity of the conclusions to the value of $\lambda$,
we also fit the data using $\lambda/a = 0.9$ GeV, which yields
an upper bound on the $16^3$ box about 50\% smaller than the $\lambda/a = 0.7$
GeV fit.
In Table \ref{tab:err_volume}, we give the results of fitting the $12^3,
16^3,$ and $20^3$ LQCD Monte Carlo results to the finite volume formula
Eq. (\ref{eq:finite_volume}). For $aE_1$ and $\tilde{f}_B$, the Table
gives the fitted infinite volume result, and an upper bound on the
finite volume term evaluated on $16^3$ and $20^3$ lattices.

\begin{table}[htb]
\centering
\renewcommand{\baselinestretch}{1.0}
\caption{Estimate of finite volume corrections to heavy-light results for
$\beta = 5.9$. For each $\kappa$, results include the fitted infinite volume
value for the ground state energy and decay constant, as well as estimated
upper bounds on the finite volume corrections on $16^3$ and $20^3$ boxes.
Unbracketed and bracketed numbers result from fitting to a Luscher asymptotic
form and to a power law ($L^{-3}$) form, respectively.}
\vspace{.1in}
\label{tab:err_volume}
\begin{tabular}{|c|c|c|c|c|c|}
\hline
\multicolumn{1}{|c|}{$\kappa$}
&\multicolumn{1}{c|}{$0.154$}
&\multicolumn{1}{c|}{$0.157$}
&\multicolumn{1}{c|}{$0.158$}
&\multicolumn{1}{c|}{$0.159$}
&\multicolumn{1}{c|}{$\kappa_c$}  \\ \hline
$aE_1(\infty)$ & .719(5) & .678(6)  & .664(7)   & .645(9) & .638(9) \\
               &[.719(7)]&[.679(9)] &[.664(11)] &[.638(13)]&[.636(13)]\\
$\Delta (aE_1)(16)$ & $\pm$ .002 & $\pm$ .002 & $\pm$ .003 & $\pm$ .004
&$\pm$.003 \\
                    &[$\pm$ .006]&[$\pm$ .008]&[$\pm$ .008]&[$\pm$
.012]&$[\pm$.011]\\
$\Delta (aE_1)(20)$ & $\pm$ .0003 & $\pm$ .0004 & $\pm$ .0004 & $\pm$
.0006&$\pm$.0005 \\
                    &[$\pm$ .003]&[$\pm$ .004]&[$\pm$ .005]&[$\pm$
.006]&[$\pm$.006]\\
$\tilde{f}_B(\infty)$  & .346(11) &  .300(11) & .283(12) & .257(14)& .246(14)\\
                       &[.348(16)]&
[.299(17)]&[.280(19)]&[.248(23)]&[.244(21)]\\
$\Delta \tilde{f}_B(16)$ & $\pm$ .004 & $\pm$ .004 & $\pm$ .004 & $\pm$
.006&$\pm$.005\\
                         &[$\pm$ .013]&[$\pm$ .014]&[$\pm$ .015]&[$\pm$
.006]&[$\pm$.005]\\
$\Delta \tilde{f}_B(20)$ & $\pm$ .0006 & $\pm$ .0007 & $\pm$ .0007 & $\pm$
.0011&$\pm$.0008 \\
                         &[$\pm$ .007]&[$\pm$ .007]&[$\pm$ .008]&[$\pm$
.011]&[$\pm$.009]\\
\hline
\end{tabular}
\end{table}
\renewcommand{\baselinestretch}{1.5}
{}From Table \ref{tab:err_volume} it is seen that, in all cases,
the estimated finite volume effect on both $aE_1$ and on $f_B$ is smaller
than our statistical error by more than a factor of two on the
$16^3$ lattice and by more than an order of magnitude on the $20^3$ lattice
at $\beta=5.9$. To determine the size of finite volume effects on the
quantities $M_{B_s}-M_{B_d}$ and $f_{B_s}/f_{B_u}$, we also need to estimate
the error on the slope
parameters $E_1^{\prime}(0)$ and $f_B^{\prime}$ in
Table ~\ref{tab:heavy_light_crit}. From the $\kappa$-dependence of the
finite volume fit parameters, we estimate
an approximate upper bound on the finite volume error for the slopes
to be $(\Delta E_1^{\prime})/E_1^{\prime}<.03$ and
$(\Delta\tilde{f}_B^{\prime})/\tilde{f}_B^{\prime}<.05$ for the $16^3$ box,
and by $(\Delta E_1^{\prime})/E_1^{\prime}<.005$ and
$(\Delta\tilde{f}_B^{\prime})/\tilde{f}_B^{\prime}<.008$ for the $20^3$ box.
Again this is about a factor of two below our statistics for $16^3$ and
entirely negligible for $20^3$.

Recently, it has been argued \cite{Fukugita} that, in intermediate ranges of
volume where the asymptotic behavior predicted by Luscher's volume formula
has not yet set in, the volume dependence might be expected to exhibit a
power law dependence of the form
\begin{equation}
\label{eq:power}
m(L) = m(\infty) + \frac{const.}{L^3}
\end{equation}
instead of the exponential falloff of Luscher's result. This power law
form is also found by the authors of Ref.\cite{Fukugita}
to fit better to their data on light hadron masses (in full QCD).
If we assume a similar power-law dependence
for the heavy-light data, we obtain extrapolated infinite volume results
and bounds on finite volume corrections which differ from
those obtained with Luscher's form. The values in Table \ref{tab:err_volume}
which are enclosed in square brackets are the results obtained by assuming
a power law dependence of the form (\ref{eq:power}). Notice that the
extrapolated
infinite volume values change very little compared with the previous analysis.
The bound on the finite volume effects at $16^3$ are somewhat larger, while
those on the $20^3$ lattice are much larger. However, in all cases, the
bound on the finite volume effect is less than the statistical error.

To estimate finite size effects for the other $\beta$
values, it is reasonable to assume approximate scaling. The box
sizes for the other $\beta$'s have been selected so that they are all
of about the same physical size as the $16^3$ box at $\beta=5.9$ (between
1.8 and 2.0 fm). Thus, we conclude that finite size effects on all
of our data is smaller than our present statistical errors.
In order to quote a systematic error on our final results (see Section 6.1)
 for $f_{B_s}/f_{B_u}$,
$M_{B_s}-M_{B_u}$ and $f_B$, we have assumed that the percentage errors for
the other $\beta$ values are the same as those obtained at $\beta=5.9$ on the
$16^3$ box.

\subsection{Extrapolation to $\kappa_{c}$}

\indent  To investigate the sensitivity of the chirally
extrapolated mass and f-values
to the fitting range in $\kappa$, we have done a detailed study of the
dependence of the results of the global (in $\kappa$) chi-square fit
(\ref{eq:chisquaredkappa}) on the $\kappa$ values chosen, for the case
$\beta=5.9$ on a 16$^{3}$ lattice. For this run, correlators were studied at
$\kappa$ values
of 0.154, 0.156, 0.157, 0.158, and 0.159 (with the critical
$\kappa_c=$0.15975).
The fits were done using a Euclidean time window $T_{<}=3, T_{>}=10$.
By taking various subsets of $\kappa$ values to perform the chiral fit (cf.
discussion at end of Section(3)), we can probe the sensitivity of our results
to the assumption of linearity of mass and f-values in $\kappa^{-1}$. The
central values obtained from the fit (together with the associated {\em
statistical} errors)
are displayed in Table \ref{tab:kappasyst}. The range of kappa
values used in the fit is indicated in the first column using the abbreviated
notation $\kappa=0.15x \rightarrow x$ (thus: $467$ indicates that the
extrapolation to $\kappa_c$ was made using $\kappa$ values 0.154, 0.156, and
0.157).

\begin{table}[htb]
\centering
\renewcommand{\baselinestretch}{1.0}
\caption{Estimate of systematic effects in chiral extrapolation.
Listed are the values at $\kappa_c$
and slopes (as a function of $\kappa^{-1}-\kappa_c^{-1}$)
of the ground state energy and decay constant
obtained from a fit to subsets of the five $\kappa$ values, .154, .156, .157,
.158, and .159. First column indicates the set of $\kappa$'s used by listing
the
last digit of each $\kappa$ included.}
\vspace{.1in}
\label{tab:kappasyst}
\begin{tabular}{|c|c|c|c|c|c|}
\hline
\multicolumn{1}{|c|}{$\kappa$ range}
&\multicolumn{1}{c|}{$E_1(0)$}
&\multicolumn{1}{c|}{$E_1^{\prime}(0)$}
&\multicolumn{1}{c|}{$\tilde{f}_B(0)$}
&\multicolumn{1}{c|}{$\tilde{f}_B^{\prime}(0)$}
&\multicolumn{1}{c|}{$\chi^2/dof$}  \\ \hline
467 & .643(10) & .319(22) & .464(29)  & .689(68) & 1.03  \\
678 & .645(10) & .303(36) & .463(29)  & .698(110) & 0.96  \\
789 & .642(13) & .332(72) & .458(36)  & .744(202) & 0.87  \\
89  & .640(15) & .385(123) & .453(41)  & .832(346) & 0.84 \\
6789 & .643(12) & .321(52) & .459(33) & .725(146) & 0.89 \\
46789 & .643(11) & .320(33) & .460(31) & .700(95) & 0.92 \\
\hline
\end{tabular}
\end{table}
\renewcommand{\baselinestretch}{1.5}
   Referring to Table \ref{tab:kappasyst}, we see that the variation in the
extrapolated ground state mass and f-value obtained by choosing three sliding
windows
of adjacent kappa values are in every case considerably smaller than the
associated statistical errors. For the mass, the central values vary by about
50 \% of the statistical error, while for the f-value the variation is
20-25 \% of the statistical error. Even totally nonoverlapping fits
(rows 467 and 89) give central values lying well within the statistical
errors. The statistical errors of course tend to
increase as we approach $\kappa_c$ ; it is more difficult to detect a
systematic trend in the central values because the dominant errors
are statistical. For the slopes
(derivatives with respect to $\kappa^{-1}$)
needed for the extraction of $B_s$ properties, the situation is
similar. Aside from the 89 fit, which gives a poor determination of the slopes,
the central values for all the subsets of $\kappa$'s are well within a standard
deviation of the full fit to all five kappa values.

  We may conclude from the preceding that, as in the case of finite volume
corrections, nonlinearities in the chiral extrapolation are not an important
source of systematic error in our results. In order to arrive at an actual
estimate
of the chiral extrapolation contribution to the total systematic error we have
taken the variation in the 3-kappa fits in Table(\ref{tab:kappasyst}) (i.e rows
labelled 467,678, and 789) which have a reasonable lever arm in $1/\kappa$, and
fairly small statistical errors, as an indication of the extrapolation error
to $\kappa_c$ (i.e. in $f_{B_u}$ and $M_{B_u}$). As we measure quite close
to $\kappa_s$ (at $\kappa=$0.157), there is effectively no extrapolation
error in the strange quark quantities. We assume that the chiral extrapolation
at $\beta$=5.9 is typical of other $\beta$ values. In this way a chiral
extrapolation part of the total systematic errors quoted in Section 6.1 can
be obtained.

\subsection{ Scale Errors}

\indent In order to quote physical values of masses and decay constants, one
must
select a particular dimensionful quantity to define the scale. In our
discussion, we have taken the values of $a^{-1}$ obtained
from the 1P-1S
charmonium splitting\cite{ElKhadra} at $\beta=5.7, 5.9,$ and $6.1$. Our choice
of $a^{-1}$ at $\beta=6.3$ is obtained by evolving from $\beta=6.1$ via
one-loop asymptotic freedom. (The same value of $a^{-1}=3.08$ is also
obtained from our value for $m_{\rho}$.)
Other possible choices for the scale-defining
parameter include string tension, rho mass, and $f_{\pi}$. Since the quoted
values of the decay constant $f_B$ include a factor of $a^{\frac{3}{2}}$,
it is particularly important to estimate the possible systematic error
in our results arising from uncertainty in the overall scale at each $\beta$
value. In Fig.~\ref{fig:scales} we have plotted the scale obtained from
$m_{\rho}$ (circles and filled circles), $f_{\pi}$ (squares), and string
tension
(diamonds and filled diamonds) relative to the scales chosen in this paper
(Table \ref{tab:scales}). Also included on the plot are points (filled squares)
obtained from lattice calculations of the deconfinement temperature $T_c$
\cite{Gottlieb}. Since the experimental value of $T_c$ is not known, these
calculations only give a relative determination of the scale at different
$\beta$'s.
(The absolute scale for these points has been chosen to be equal to that in
Table \ref{tab:scales} at $\beta=6.3$, which corresponds to a deconfinement
temperature of $kT_c = 264$ MeV.)
The values for
$m_{\rho}$ are from  GF11 \cite{Weingarten} (filled circles) and from
our data (Table \ref{tab:light_meson})(circles), while the
string tension is a combination of our results at 5.7, 5.9, and 6.1 (diamonds)
and those of Ref.~\cite{Bali} at 6.0, 6.2, and 6.4 (filled diamonds).
For the latter points, the
charmonium scales were estimated from Table \ref{tab:scales} by linear
interpolation in $\ln\,a$. The values for $f_{\pi}$ are taken from
Table \ref{tab:light_meson}.

\begin{figure}[htp]
\label{fig:scales}
\end{figure}

The trend exhibited by the data in Fig. \ref{fig:scales} indicates a
significant scale discrepancy in the range $\beta = 5.7$ to 6.0, with the
$m_{\rho}$ scale being about 10-20\% higher and the string-tension scale
about 10-15\% lower than charmonium. For $\beta\geq 6.2$, the scales appear
to converge to much better agreement, with deviations of $\leq 5\%$. This
suggests that much of the discrepancy at lower $\beta$ is due to finite lattice
spacing effects, as opposed to being an effect of the quenched approximation.
(Discrepancies which do not go away in the scaling limit can be ascribed to
the neglect of closed quark loops.) The data shown in Fig. \ref{tab:scales}
illustrates that, over the entire range of $\beta$, the charmonium scale
differs little from a weighted average of the other choices. This provides some
additional confidence in our choice of scales. To estimate the scale
error on our heavy-light results, we have used the charmonium scale
errors quoted in Ref.
\cite{ElKhadra} (which include both statistics and systematics).
For $\beta = 6.3$ we have taken a conservative scale error estimate of $5\%$,
based on the spread of values shown in Fig. \ref{fig:scales}.

  Our final results for $M_{B_s}-M_{B_u}$ and $f_{B_u}$ ($f_{B_s}/f_{B_u}$ is
dimensionless) quoted below are therefore subject to a 5\% and 7\% error
(resp.),
assuming that continuum extrapolated objects are determined primarily by the
larger $\beta$ values where the scale discrepancy is small. The larger error
for $f_B$ arises from the fact that the quantity computed on the lattice
scales like $a^{3/2}$.


\subsection{Extrapolation to the continuum}

By far the largest systematic error in our
calculations arises in the extrapolation
of the $f_{B}$ results to zero lattice spacing.
In comparison, the systematic errors
incurred from working on a finite volume lattice,  at finite light quark mass,
or even (very probably) the neglect of quark loops are negligible.
The difficulties
here are both intrinsic and practical. On the one hand, the detailed form of
the
lattice spacing dependence of lattice quantities is generally rather
complicated
(involving logarithmic as well as power dependence on the lattice spacing), in
contrast to the relatively well understood structure of the
chiral or finite volume extrapolations.
On the other hand, reduction of the lattice spacing by a factor of 2 requires
increasing the lattice volume 16-fold (if we maintain fixed physical space-time
volume).

\begin{figure}[htp]
\label{fig:fb_fpi_vs_a}
\end{figure}

 These issues are particularly important in the case of heavy meson decay
constants.
We find that the lattice spacing dependence for $f_{B}$ in the static limit is
considerably stronger than for $f_{\pi}$. This is illustrated
in Fig. \ref{fig:fb_fpi_vs_a}. Although our calculations have led to reasonably
precise results at finite a, quoting a systematic error on the continuum
extrapolated
result clearly requires an investigation of the variation induced by
alternative
fitting procedures.

\begin{table}[htb]
\centering
\renewcommand{\baselinestretch}{1.0}
\caption{Comparison of continuum values for a linear vs. a quadratic fit
to the $a$-dependence of physical quantities. Values shown are the fitted
$a=0$ values of $f_B, M_{B_s}-M_{B_u},$ and $f_{B_s}/f_{B_u}$, and the
$\chi^2$ per degree of freedom for each fit. Errors shown are extrapolated
statistical errors.}
\vspace{.1in}
\label{tab:cont_extrap}
\begin{tabular}{|c|c|c|c|c|c|c|}
\hline
 fit & $f_B (MeV)$ & $\chi^2/dof$ & $M_{B_s}-M_{B_u}$ & $\chi^2/dof$ &
 $f_{B_s}/f_{B_u}$ & $\chi^2/dof$ \\ \hline
$c_0+c_1a$  & 188(23) & .46/2 & 86(12) & 2.4/2 & 1.22(4) & 2.4/2  \\
$c_0+c_1a^2$& 214(13) & .40/2 & 80(7)  & 2.3/2 & 1.21(2) & 2.2/2  \\
\hline
\end{tabular}
\end{table}
\renewcommand{\baselinestretch}{1.5}
We have investigated this variation by fitting the $f_B$ results at finite
$\beta$ to various functional forms for the finite $a$ correction. In
Table \ref{tab:cont_extrap} we show the extrapolated continnum result for $f_B$
and the
$\chi^2$ per degree of freedom for the alternatives that the finite lattice
spacing
correction is of order $a$ and of order $a^2$.
The linear and square fits are, explicitly,
\begin{eqnarray}
f_B(a) & = & 188(23)\left(1 + .51(19)a\right)\,\,\mev \\
f_B(a) & = & 214(13)\left(1 + \left[.60(11)a\right]^2\right)\,\,\mev
\end{eqnarray}
where $a$ is in $GeV^{-1}$.
For $M_{B_s}-M_{B_u}$, the fits are
\begin{eqnarray}
M_{B_s}-M_{B_u}(a) & = & 85.8(11.7)\left(1 - .24(22)a\right)\,\,\mev \\
M_{B_s}-M_{B_u}(a) & = & 80.1(6.7)\left(1 -
\left[.47(40)a\right]^2\right)\,\,\mev
\end{eqnarray}
and
\begin{eqnarray}
f_{B_s}/f_{B_u}(a) & = & 1.216(41)\left(1 - .02(6)a\right) \\
f_{B_s}/f_{B_u}(a) & = & 1.213(23)\left(1 - \left[.16(31)a\right]^2\right)
\end{eqnarray}
These fits are shown with the data in Fig. \ref{fig:fb_fpi_vs_a}.
It is clear from this graph that it would be very difficult to
distinguish between these two possibilities by calculations in the range
$\beta = 5.7-6.3$, even with improved statistics and additional $\beta$
values.
In the absence of more precise data at much larger $\beta$ or
a complete control of all $O(a)$ lattice effects,
we will take the variation of the extrapolated results in
Table \ref{tab:cont_extrap} as an estimate of the systematic error associated
with extrapolating to $a=0$.
In our final result for $f_B$, we quote the continuum
value obtained from the linear fit with asymmetric errors of +26 and -0
to allow for this extrapolation uncertainty.


\newpage
\section{Physical Results and Discussion}

\subsection{Final results}

  In this final section, we collect our results for $f_{B_s}/f_{B_u}$,
$M_{B_s}-M_{B_u}$, and $f_{B_u}$ at each lattice spacing calculated
and then present the results extrapolated to the continuum.
The summary of results
for each of the four lattice
spacings studied are collected in Table \ref{tab:final_results}.
A comparison with other recently reported
results follows immediately. In all cases the first error quoted is a
statistical error obtained by the jackknife procedure described previously.

\begin{table}[htp]
\renewcommand{\baselinestretch}{1.0}
\centering
\caption{Final results at fixed $\beta$ (lattice spacing)
for $f_B, M_{Bs}-M_{Bu},$ and $f_{Bs}/f_{Bu}$ in the
static approximation. The first error is statistical and the second is
systematic.}
\vspace{.1in}
\label{tab:final_results}
\begin{tabular}{|c|c|c|c|}
\hline
\multicolumn{1}{|c|}{$\beta$}
&\multicolumn{1}{c|}{$f_B (\mev)$}
&\multicolumn{1}{c|}{$M_{Bs}-M_{Bu} (\mev)$}
&\multicolumn{1}{c|}{ $f_{Bs}/f_{Bu}$} \\ \hline
5.7  & $271\pm 13\pm 20$ & $66\pm 7\pm 6$ &$1.181\pm .030\pm .012$\\
5.9  & $241\pm 13\pm 13$ & $74\pm 5\pm 6$ &$1.211\pm .018\pm .014$\\
6.1  & $215\pm 21\pm 14$ & $87\pm 9\pm 7 $&$1.226\pm .027\pm .016$\\
6.3  & $225\pm 17\pm 14$ & $68\pm 10\pm 5$ &$1.172\pm .031\pm .011$\\
\hline
\end{tabular}
\end{table}
\renewcommand{\baselinestretch}{1.5}

 The source of systematic errors varies somewhat with the quantity being
computed. For $f_{B_s}/f_{B_u}$, the uncertainty in the scale cancels, as the
ratio is a dimensionless quantity. Thus, for this quantity, the quoted
systematic error includes finite volume and chiral extrapolation errors.
The lattice spacing dependence of this quantity (see Fig(7)) is very small,
so we have not included a continuum extrapolation error. The result is
\begin{equation}
   f_{B_s}/f_{B_u}= 1.216  \pm 0.041 (\rm stat) \pm 0.016 (\rm syst)
\end{equation}

  For the mass splitting $M_{B_s}-M_{B_u}$ the systematic errors include
finite volume effects, the chiral extrapolation (to determine $M_{B_u}$),
and an estimate of the scale error. For the continuum extrapolation, we quote
the result obtained from the linear fit in $a$, and take the difference
between the two fits shown in Table 11 as an estimate of our extrapolation
error ($+0$ to $-6$). Combining the extrapolation error with our other
systematic errors, we obtain the result
\begin{equation}
   M_{B_s}-M_{B_u}= 86 \pm 12 (\rm stat) ~^{+7}_{-9} (\rm syst)\,\, MeV.
\end{equation}

 Finally, for $f_B$ itself, there is, in addition to the usual finite volume
and chiral extrapolation errors, a substantial (not necessarily linear)
lattice spacing dependence, as well as the possibility of
sizable renormalization
corrections to $Z$ at the two and higher loop level. These additional
systematic errors are quoted separately in our final result:
\begin{equation}
  f_{B_u} = 188 \pm 23 (\rm stat) \pm 15 (\rm syst) ~^{+26}_{-0} (\rm extrap)
\pm 14 (\rm renorm)\,\,MeV.
\end{equation}

\subsection{Comparison with other results}

There have been several recently reported studies of the heavy-light
meson system in the static limit of quenched LQCD. In this Section
we will assess the results reported here in comparison with these
other studies. We will consider four quantities, $E_1$, the heavy-light
ground state energy, the $M_{B_s}-M_{B_u}$ mass splitting,
$f_B$ the ground-state decay constant, and the ratio $f_{B_s}/f_{B_u}$.
Even before
the present study, there has been some apparent disagreement among different
groups over the size of $f_B$. Some of these discrepancies can be
traced to different choices of $\beta$ and $\kappa$, different definitions
of the QCD length scale, and different evaluations of the perturbative
renormalization constant $Z$. Since our data has provided a more accurate
determination of the dependence on $\kappa$ and $\beta$, we are able to
interpolate our data in both variables and make a direct comparison with
other groups.

The focus of this subsection is on the lattice measurement of the ground state
energy eigenvalue $E_1$ and its matrix element with the unrenormalized
local axial current operator on the lattice, $\tilde{f}_B = v^N_1$ (c.f.
Eq.(\ref{eq:chisquared})).
These are the quantities that are extracted directly from the lattice
heavy-light meson propagators and, for given $\beta$ and $\kappa$, are
independent of the choices for length scale and renormalization constant.

\begin{figure}[tbp]
\label{fig:massplot}
\end{figure}

First consider the ground state energy $E_1$. In  Section 4.1 it was
shown that the $\kappa$ and $a$ dependence of $E_1$ is most easily described
by introducing the naive quark mass $m_q = (2\kappa a)^{-1}-(2\kappa_c
a)^{-1}$.
The values of
$\kappa_c$ for $\beta = 5.74$ and 6.26 are taken from \cite{Wuppertal} while
that at $\beta=6.0$ is taken from \cite{BLS}. The $\kappa_c$ values for
$\beta=5.7, 5.9, 6.1,$ and $6.3$ are from our own data. The values for the
scale
$a^{-1}$ are subject to somewhat more uncertainty. For 5.7, 5.9, 6.1, and 6.3,
the scales we have used are given in
Table \ref{tab:scales}. In order to have a reasonably
self-consistent set of scales, the remaining values for $a^{-1}$ are obtained
from those at 5.7, 5.9, and 6.1 by a simple linear interpolation or
extrapolation in $\ln a$. Note that we are only using the value of
$a^{-1}$ here to define the naive quark mass $m_q$. The numerical values of
$E_1$ and $\tilde{f}_B$ are obtained directly from the lattice propagators
and do not depend on choice of scale. Thus, our comparison of data is
insensitive to an overall, uniform change of scales. The values for the
ground state energy $E_1$ (in lattice units), extrapolated to $\kappa_c$,
 from the various studies are plotted in
Fig.\ref{fig:massplot}.
We conclude from these plots that all the data are in reasonable agreement,
both on the magnitude of $E_1$ and on its  $a$ dependence.
This is not surprising since, for any reasonably well-chosen smearing function,
the value of $E_1$ is obtained fairly unambiguously from the smeared-smeared
correlator. Our multistate analysis provides a value of $E_1$ with considerably
smaller errors than a single-channel analysis (note the much smaller error
bars for the ACPMAPS results), but the results are completely
consistent with previous calculations.

  Results from various groups for the $M_{B_s}-M_{B_u}$ splitting are compared
with ours in Fig ~\ref{fig:splitplot}. Again, the accurate determination of
the slope of the ground state energy with respect to $1/\kappa$ is the key
to the much smaller error bars shown for the ACPMAPS results. However, the
results
are basically consistent within errors. For this quantity, the lattice spacing
dependence is very mild.
\begin{figure}[htp]
\label{fig:splitplot}
\end{figure}

Results from various groups for the ratio of decay constants
$f_{B_s}/f_{B_u}$ are compared
with ours in Fig ~\ref{fig:fbsplot}.
\begin{figure}[thp]
\label{fig:fbsplot}
\end{figure}

\begin{figure}
\label{fig:fbplot}
\end{figure}

In contrast with the situation for $E_1$ or $M_{B_s}-M_{B_u}$,
there is significant disagreement
on the value of $\tilde{f}_B$ among the various studies, even after correcting
for the different values of $\kappa$ and $\beta$.
Results from references
\cite{RomeSoton,Rome,BLS,Wuppertal,Kentucky} are compared with
our data in figure~\ref{fig:fbplot}. This is a plot of ${\tilde f}_B
a^{-3/2}$ against lattice spacing $a$, so that we can compare \lq raw'
lattice results without renormalisation factors.
The results of~\cite{BLS} (at $\beta=6.3$)
and~\cite{Kentucky} are in good agreement with ours,
while~\cite{RomeSoton,Rome,Wuppertal} report substantially
larger values.

We believe that the discrepancies can be traced primarily to the
incomplete isolation of the ground state with the sources chosen.
Observe in particular that the result in \cite{Kentucky}, where a
variational method is used to isolate the ground state, is in good
agreement with us.  Sensitivity of the apparent value of $f_B$ to the
form of the source has been emphasized in \cite{Hashimoto}.

\subsection{Future Studies}

We have presented results for the decay constant $f_B$ and for masses of
low-lying heavy-light states in the static approximation.
The analysis
procedure introduces several
improvements over previous smearing methods. First, the success of the
RQM in reproducing the measured lattice wave functions is exploited by
using the RQM to construct not only an accurate ground state wave function,
but also a set of orthonormal excited state smearing functions. Second,
we make full use of the information contained in the
matrix of smeared-smeared and
smeared-local correlators, including both ground state and excited state
smearing functions at each end. Our method provides much greater
control over systematic errors from higher state contamination, because
of the fact that the source smearing functions are tuned directly to the
lattice wave functions, without regard to the behavior of the effective
mass plots. The appearance of long plateaus in the SS and SL plots at
the same value of effective mass is thus strong evidence that the systematic
error from higher states has been largely eliminated.
We will report the application of our methods for heavy-light mesons
to the spectrum of radial and orbital excitations
for heavy-light systems and the B parameter in forthcoming papers.

We expect to be
able to improve the accuracy of the present results for the
$M_{B_s}- M_{B_u}$ and $f_{B_s}/f_{B_u}$ by using
larger ensembles.
There are two other sources of systematic
uncertainty in our results for $f_B$.
Use of the Wilson action for the light quarks
implies lattice spacing corrections in O(a) and the large one loop
renormalization for the axial current suggests that the two
loop correction may be sizable. Study of these effects will be
required to substantially improve the error on $f_B$.


\appendix
\newpage
\section{Light Hadron Results}\label{appen:light}


In this Appendix we report the results for the light quark sector
which we have used to set the parameters for the heavy-light meson properties.
First consider the
light-light pseudoscalar meson mass as a function of the light quark
hopping parameters. This serves to determine the critical hopping parameter
at each value of $\beta$ studied. Using the physical $K$ mass, we also
establish the correct hopping parameter for the strange quark. (The small
effect from the nonzero up or down quark mass has also been included in our
results.) In addition to determining the light quark hopping parameters
from the pseudoscalar meson masses, the meson propagators have also been
analyzed to obtain the values of $f_{\pi}$ and $m_{\rho}$. All of the results
discussed here for the light hadron parameters have been extracted from local
$q\bar{q}$ operators. An analysis of light meson parameters
using smeared sources in Coulomb gauge
is in progress and will be reported elsewhere.

Let the hopping parameters for the two valence quarks in the light meson
be denoted by $\kappa_1$ and $\kappa_2$, and denote the pseudoscalar meson
mass by $m_P$. All of our data is consistent with a linear dependence of
$m_P^2$ on $\kappa_1^{-1}$ and $\kappa_2^{-1}$.
The results can be expressed in terms of
the parameters $C$ and $\kappa_c$ of the linear fit
\begin{equation}
\label{eq:chiral_masses}
(m_Pa)^2 = C\left(\kappa_1^{-1} + \kappa_2^{-1} - 2\kappa_c^{-1}\right)
\end{equation}
The results for the four $\beta$ values studied are given in Table
\ref{tab:kappa_c}. Using the scales in Table \ref{tab:scales}, the naive
quark masses, defined by
\begin{equation}
\label{eq:quark_mass}
m_q = \frac{1}{2\kappa a} - \frac{1}{2\kappa_c a}
\end{equation}
are also listed in Table \ref{tab:kappa_c}. (Here, up and down quarks
are taken to be degenerate in mass.)
A plot of $m_P^2$ vs. quark mass (defined in Eq. (\ref{eq:quark_mass})) is
shown in Fig. \ref{fig:pionmass_squared}.
For $\beta=5.7, 6.1$, and $6.3$ the values of $m_Pa$ were obtained using
equal quark masses, $\kappa_1=\kappa_2$. For $\beta=5.9$ several combinations
of unequal quark mass values (open circles in Fig.\ref{fig:pionmass_squared})
were used to check the validity of Eq.(\ref{eq:chiral_masses}).
For $\beta=5.9$, in addition to the main run on a $16^3\times32$
lattice, in order to investigate finite volume effects, we have also
carried out Monte Carlo runs on $12^3\times24$ and $20^3\times40$ lattices.
The results labeled $5.9(\infty)$ are the infinite volume values obtained by
fitting all three box sizes to the functional form derived by Luscher, as
discussed in Sec. 5.2.

\begin{figure}[htp]
\label{fig:pionmass_squared}
\end{figure}

\begin{table}[htb]
\centering
\renewcommand{\baselinestretch}{1.0}
\caption{Light quark hopping parameters obtained from light pseudoscalar meson
mass measurements. Values are shown for each run, as well as the
infinite volume extrapolated value at $\beta=5.9$. C is the slope of the
pseudoscalar (mass)$^2$ as a function of $\kappa^{-1}$. The last two columns
give the strange and up quark masses $\times 2a$. Errors are statistical.}
\vspace{.1in}
\label{tab:kappa_c}
\begin{tabular}{|c|c|c|c|c|c|c|}
\hline
\multicolumn{1}{|c|}{$Run$}
&\multicolumn{1}{c|}{$\beta$}
&\multicolumn{1}{c|}{ $\kappa_c$}
&\multicolumn{1}{c|}{ $C$}
&\multicolumn{1}{c|}{ $\kappa_s^{-1}-\kappa_c^{-1}$}
&\multicolumn{1}{c|}{ $\kappa_u^{-1}-\kappa_c^{-1}$} \\ \hline
b & 5.7  & .16914(10)  & .703(10) & .2533(36) & .01024(15)\\
e & 5.9  & .15972(14)  &  .615(14)  & .1209(27) & .00489(11)\\
c & 5.9  & .15975(6)  &  .609(9)  & .1221(18) & .00494(7)\\
f & 5.9  & .15981(4)  &  .591(8)  & .1258(17) & .00509(7)\\
  & 5.9($\infty$)& .15980(4) & .597(7)  & .1245(14) & .00504(6)\\
d & 6.1  & .15496(3)  &  .480(11)  & .0832(19) & .00336(8)\\
g & 6.3  & .15178(4)  &  .395(10) & .0613(15) & .00248(6)\\
\hline
\end{tabular}
\end{table}
\renewcommand{\baselinestretch}{1.5}

In Table \ref{tab:kappa_c}, the values for $C$ and $\kappa_c$ are
independent of the scale chosen for $a^{-1}$. The remaining columns
are computed using the scales in Table \ref{tab:scales}. The errors
quoted in Table \ref{tab:kappa_c} are statistical only, and do not
include the uncertainty associated with the choice of $a^{-1}$.  The
uncertainty arising from the scale determination, along with other
sources of error, will be discussed in Section 6.

\begin{table}[p]
\label{tab:light_meson}
\centering
\renewcommand{\baselinestretch}{1.0}
\caption{Results for light-light mesons. Meson propagators were fit to a single
exponential over time window $\Delta T$. Result for $f_{\pi}$ includes a
tadpole
improved renormalization factor, computed in Ref.[20].}
\vspace{.1in}
\begin{tabular}{|c|c|c|c|c|c|}
\hline
$\beta$ & $\Delta T$ & $\kappa$ & $m_{\pi}a$ & $m_{\rho}a$ & $f_{\pi}a$ \\
\hline
5.7  &6-10& .161  & .649(3)  & .787(2) & .145(3)\\
     && .165  & .456(5)  & .675(5) & .133(4)\\
     && .1667 & .351(8)  & .629(15)& .115(6) \\
     && .168  & .237(13) & .586(44)& .102(11)\\
     && $\kappa_c$ & 0   & .566(11)& .108(14)\\
5.9  &8-12& .154  & .527(2)  & .619(2) & .104(2)\\
     && .156  & .426(3)  & .546(3) & .093(2)\\
     && .157  & .360(2)  & .513(3) & .086(2)\\
     && .158  & .288(3)  & .479(5)& .078(2)\\
     && .159  & .195(4) & .444(12)& .072(4)\\
     && $\kappa_c$ & 0 & .418(4) & .067(2)\\
6.1  &10-16& .151  & .409(3)  & .482(3) & .081(2)\\
     && .153  & .276(4)  & .401(4) & .069(2)\\
     && .154  & .196(3)  & .361(8) & .061(2)\\
     && .1545 & .137(5)  & .341(19)& .056(4)\\
     && $\kappa_c$ & 0   & .324(8)& .055(2)\\
6.3  &12-18& .1500 & .249(3)  & .331(3) & .054(2)\\
     && .1510 & .160(6)  & .283(5) & .045(4)\\
     && .1513 & .126(6)  & .269(7) & .041(4)\\
     && .1515 & .099(5)  & .260(11)& .038(11)\\
     && $\kappa_c$ & 0   & .246(8) & .037(3)\\
\hline
\end{tabular}
\end{table}

\begin{table}[p]
\label{tab:light_meson_vol}
\centering
\renewcommand{\baselinestretch}{1.0}
\caption{Light-light meson results at $\beta=5.9$ for lattice sizes
$12^3\times 24 (e), 16^3\times 32 (c)$, and $20^3\times 40 (f)$. Result
for $f_{\pi}$ includes the tadpole improved renormalization factor, computed
in Ref. [20].}
\vspace{.1in}
\begin{tabular}{|c|c|c|c|c|c|}
\hline
\multicolumn{1}{|c|}{$\beta$}
&\multicolumn{1}{c|}{ $\Delta T$}
&\multicolumn{1}{c|}{ $\kappa$}
&\multicolumn{1}{c|}{ $m_{\pi}a$}
&\multicolumn{1}{c|}{ $m_{\rho}a$}
&\multicolumn{1}{c|}{ $f_{\pi}a$} \\ \hline
5.9(e)&8-12& .154  & .535(4)  & .618(4) & .105(3)\\
     && .157  & .364(4)  & .505(7) & .083(4)\\
     && .158  & .306(9)  & .475(15) & .080(4)\\
     && .159  & .198(33)  & .375(77)& .075(17)\\
     && $\kappa_c$ & 0 & .403(14) & .067(5)\\
5.9(c)&8-12& .154  & .534(3)  & .622(2) & .106(3)\\
     && .157  & .364(4)  & .508(5) & .087(3)\\
     && .158  & .292(4)  & .469(8)& .079(4)\\
     && .159  & .188(8) & .440(27)& .067(8)\\
     && $\kappa_c$ & 0 & .407(8) & .067(4)\\
5.9(f)&8-12& .154  & .525(2)  & .617(2) & .103(2)\\
     && .157  & .358(3)  & .515(4) & .085(2)\\
     && .158  & .288(3)  & .482(5) & .078(3)\\
     && .159  & .196(4)  & .442(12)& .073(4)\\
     && $\kappa_c$ & 0 & .423(5) & .068(2)\\
\hline
\end{tabular}
\end{table}
\renewcommand{\baselinestretch}{1.5}
\renewcommand{\baselinestretch}{1.5}
Although they are not used in the body of the paper, we also
briefly discuss our results for the rho mass and the pion
decay constant. We again emphasize that all of the results discussed here
are obtained from light quark propagators with $\delta$-function
sources. The results are thus subject to possible systematic errors
from higher state contamination. This is not a problem in measuring the
pion and kaon masses discussed above, but
it becomes more of a difficulty for $m_{\rho}$ and $f_{\pi}$ measurements.
In order to determine $f_{\pi}$, one must calculate both the
propagator with a pseudoscalar ($\bar{q}\gamma_5 q$) source at each end
(PP), and the propagator with a pseudoscalar source at one end and an
axial vector ($\bar{q}\gamma_5\gamma_0 q$) source at the other end (PA).
(Note: One may also use the (AA) propagator with an axial vector source
at both ends. However, since the vacuum to one pion matrix element of
the axial vector source contains an explicit factor of $m_{\pi}$, the
pion pole residue in the (AA) propagator vanishes more rapidly than that
of the (PA) propagator in the chiral limit, making it more difficult to
measure accurately.) The (PP) and (PA) propagators are fit to a single
exponential in the time ranges shown.
Table \ref{tab:light_meson} summarizes our results for light hadron parameters.
Again, the errors quoted are purely statistical. Some systematic errors are
expected particularly for $f_{\pi}$, for which a stable mass plateau in
the (PA) propagator with the same mass as the (PP) propagator
was not generally achieved. An attempt was made to compensate for this
by fitting the (PA) propagator with the mass fixed to be equal to that
of the (PP) propagator. It is clear from these results that a much
better determination of $f_{\pi}$ from our data will be possible when smeared
operators are employed. The rho propagators exhibit reasonable
plateaus in the time intervals shown in Table \ref{tab:light_meson}, but a
study of the variation of the results with different $\Delta T$ windows
indicates that a systematic error of from 1 to 2 times the statistical
errors cannot be ruled out. The values of $f_{\pi}a$ given in
Table \ref{tab:light_meson} include the perturbative renormalization
constants computed in Ref. \cite{LepMack}. For $\beta=5.9$, the results
in Table \ref{tab:light_meson} are the infinite volume extrapolated results
from the three Monte Carlo runs on $12^3$ (e), $16^3$ (c), and $20^3$ (f) boxes
(except for $\kappa=.156$ which was done only on the $16^3$ box). The results
on each size box are listed separately in Table \ref{tab:light_meson_vol}.
In the main analysis of this paper, the only
light meson parameters we will use are the hopping parameters determined
from the pseudoscalar masses. Since the (PP) propagator always exhibits
a stable mass plateau, these parameters are well-determined and should
be relatively free of systematic error from higher states.


\newpage
\section{Multistate Extraction of Meson Wavefunctions}\label{appen:wf}

  A particularly graphic illustration of the power of the smearing
technique in reducing the contamination of higher states is obtained
by examining the time development of the Coulomb gauge Bethe-Salpeter
wavefunction of a static-light meson beginning with either (a) the RQM
smeared source defined in equation~(\ref{eq:smear}), or (b) a cube smeared
source \cite{Rome}.
In the first case, we extract
\begin{equation}
  \Psi_{BS}(R,T)\equiv <0|q(R,T)\bar{Q}(0,T)|\Phi^{(a)},0>
\end{equation}
for the ground state ($a$=1) at small Euclidean times.  We have done the
comparison for the case $\beta=$5.9, $\kappa=$0.159, on a 16$^{3}$
lattice.  In Fig.\ref{fig:1Srqm} the evolution of the wavefunction using
a source smeared with the ground state wavefunction of the RQM (with
constituent mass $\mu$=0.12) is shown for Euclidean times $T$=1,2 and 4.
It is apparent that the wavefunction has reached its asymptotic value
to very good accuracy already at time slice 2, with little further
change at $T$=4 (in fact, the overlap of the wavefunction at $T$=2
with that at $T$=4 is 0.9986!).  In Fig.\ref{fig:1Scube}
the corresponding evolution (again for Euclidean times $T$=1,2 and 4)
is shown for a source smeared over a cube of width 7 lattice spacings.
Here the convergence is much slower, with the pointwise convergence
near the origin particularly tardy.

  Although the use of a single smearing function obtained
from the RQM is adequate to the
task of extracting the ground-state Bethe-Salpeter
wavefunction, even the improved smearing
given by the RQM is not sufficient if we wish
 to do the same for the higher excited states
in a given channel. In Section 4 we showed how to define
optimized smeared states $\mid\hat{\Phi}^{A}>$ in which the
admixture of all but one (the $A$'th) of the first
 $M$ meson states in a given channel is tuned to zero.
For example, taking $M$=3 at $\beta$=5.9, $\kappa$=0.159 on a 16$^{3}$ lattice,
one finds that the
choice of smearing function (cf. Sect 3.2,Eq~(\ref{eq:antisymm}))
\begin{equation}
\label{eq:optsmear}
  \mid\hat{\Phi}^{(2)}>=0.03\mid \Phi^{(1)}>+0.82\mid \Phi^{(2)}>
   +0.57\mid \Phi^{(3)}>
\end{equation}
produces an optimized first excited state
in the sense that admixtures of the ground and second
excited state are tuned out (based on a fit of the form ~(\ref{eq:chisquared}))
with
$T_{<}$=2,$T_{>}$=7). The use of
such an optimized smearing is crucial if we wish to extract the correct
Bethe-Salpeter wavefunction
of the first radial excited state near the origin. Any sizable admixture of the
exact meson
ground-state will otherwise dominate the small $r$ region of the wavefunction
at
large Euclidean time,
{\em before} the higher (2nd, 3rd,etc) states have decayed away.
In Figure \ref{fig:2Srqm} we show this
phenomenon with a Bethe-Salpeter wavefunction $\Psi_{BS}(R,T)$ defined as
\begin{equation}
\label{eq:nonopt}
   \Psi_{BS}(R,T) \equiv <0\mid q(\vec{R},T)\bar{Q}(0,T)\mid \Phi^{(2)},0>
\end{equation}
The wavefunction (renormalized to unit norm)
is plotted for Euclidean times $T$=1,2,4, and 6.
There is a steady upward drift of
the wavefunction at the origin as $T$ increase to 4,
but by time slice 6 the influence of the
ground state is clearly apparent as the latter
begins to dominate the evolved meson state. It would
clearly be very difficult to draw any firm conclusions
about the behavior of the excited state
wavefunction close to the origin from these measurements.

  On the other hand, using  the optimized smearing found above
{}~(\ref{eq:optsmear}),
and computing
\begin{equation}
  \Psi_{BS}^{\rm opt}(R,T) \equiv <0\mid
q(\vec{R},T)\bar{Q}(0,T)\mid\hat{\Phi}^{(2)},0>
\end{equation}
one finds (Fig.\ref{fig:2Sopt} shows the optimized wavefunction for times
$T$=1,2,3)
a rapid convergence to an asymptotic shape by the
 {\em third} time slice, giving a value for the
wavefunction at the origin $\simeq$0.44, as compared to
a maximal value $\simeq$0.36 obtained at $T$=4
from Eq~(\ref{eq:nonopt}) before convergence is lost. Eventually, of course,
 the ground state will dominate in
this case also, but by using the optimized state,
we correctly extract the exact Bethe-Salpeter
wavefunction for the excited state before the ground state
contamination has a chance to become
sizable.

\begin{figure}[htp]
\label{fig:1Srqm}
\end{figure}

\begin{figure}[htp]
\label{fig:1Scube}
\end{figure}

\begin{figure}[htp]
\label{fig:2Srqm}
\end{figure}

\begin{figure}[htp]
\label{fig:2Sopt}
\end{figure}


{\bf \hspace{-.5in} ACKNOWLEDGEMENTS}


The authors wish to acknowledge the contributions of
Aida El-Khadra to this work.
We have also thank Andreas Kronfeld and Paul Mackenzie for valuable
discussions.
JMF thanks the Nuffield Foundation for
support under the scheme of Awards for Newly Appointed
Science Lecturers.
BRH was supported in part by the Department of Energy
under Grant No.~DE--FG03--91ER 40662, Task C.  HBT was supported in part by
the Department of Energy under Grant No.~DE-AS05-89ER 40518.
AD was supported in part by the National Science Foundation
under Grant No. PHY-90-24764.
This work was performed using the ACPMAPS computer
at the Fermi National Accelerator Laboratory, which is operated by
Universities Research Association, Inc., under contract DE-AC02-76CHO3000.
\newpage


{\Large Figure Captions}
\vspace*{1in}
\newcounter{figpage}
\begin{list}%
{Figure~{\arabic{figpage}}:}{\usecounter{figpage}}

\item Static potential calculated in Coulomb gauge at
$\beta = 5.7, 5.9, 6.1$ and, $6.3$ on lattices of size $~12^3\times 24$,
$16^3\times 32$, $24^3\times 48$, and $32^3\times 48$, respectively. Errors
shown are statistical.

\item Smeared-smeared (S-S) and
smeared local (S-L) effective masses for
$\beta = 5.7, 12^3\times 24$, and $\kappa = .161, .165, .1667, and .168$. Solid
line is the ground state energy extracted from a 2-state fit. Smearing
functions are optimized combinations of RQM wavefunctions, as described in
the text.

\item Smeared-smeared (S-S) and
smeared local (S-L) effective masses for
$\beta = 5.9, 12^3\times 24$, and $\kappa = .154, .157, .158, and .159$. Solid
line is the ground state energy extracted from a 2-state fit. Smearing
functions are optimized combinations of RQM wavefunctions, as described in
the text.

\item Smeared-smeared (S-S) and
smeared local (S-L) effective masses for
$\beta = 5.9, 16^3\times 32$, and $\kappa = .154, .157, .158, and .159$. Solid
line is the ground state energy extracted from a 2-state fit. Smearing
functions are optimized combinations of RQM wavefunctions, as described in
the text.

\item Smeared-smeared (S-S) and
smeared local (S-L) effective masses for
$\beta = 5.9, 20^3\times 40$, and $\kappa = .154, .157, .158, and .159$. Solid
line is the ground state energy extracted from a 2-state fit. Smearing
functions are optimized combinations of RQM wavefunctions, as described in
the text.

\item Smeared-smeared (S-S) and
smeared local (S-L) effective masses for
$\beta = 6.1, 24^3\times 48$, and $\kappa = .151, .153, .154, and .1545$. Solid
line is the ground state energy extracted from a 2-state fit. Smearing
functions are optimized combinations of RQM wavefunctions, as described in
the text.

\item Smeared-smeared (S-S) and
smeared local (S-L) effective masses for
$\beta = 6.3, 32^3\times 48$, and $\kappa = .1500, .1500 .1513, and .1515$.
Solid line is the ground state energy extracted from a 2-state fit. Smearing
functions are optimized combinations of RQM wavefunctions, as described in
the text.

\item Heavy-light ground-state energy $aE_1$ vs. bare quark mass for
$\beta = 5.7, 5.9, 6.1$, and $6.3$
Each data point is the ground state energy extracted from a two-state fit.
Error bars are statistical only.
Solid lines are obtained from
a simultaneous two-state fit to all kappa
values for a given $\beta$ as described in the text.

\item $\tilde{f}_B$ as a function of bare quark mass
for the four runs $\beta=5.7, 5.9, 6.1$, and $6.3$.
Data points are decay constants extracted from a two-state fit.
Error bars are statistical only.
Solid lines are obtained from
a simultaneous fit to all kappa values for a given beta as described
in the text.

\item Heavy-light ground-state energy
at $\kappa=\kappa_c$ vs. lattice
spacing for $\beta = 5.7, 5.9,.6.1,$ and$6.3$. Solid line represents a
minumum-$\chi^2$ linear fit to the four data points.
\item Figure 11:~$M_{B_s}-M_{B_u}$ vs. lattice spacing.
Solid line is the best linear
fit. Dashed line is a quadratic fit ($a^2$) used to estimate systematic
error in $a\rightarrow 0$ extrapolation. (See Section 5.5).

\item $f_{Bs}/f_{Bu}$ vs. lattice spacing.
Solid line represents
the best linear ($a$) fit.
Dashed line is a quadratic fit ($a^2$) used to estimate systematic
error in $a\rightarrow 0$ extrapolation. (See Section 5.5).

\item Time-window dependence of effective mass
for 1-state fits compared
with 2-state fits for $\beta=5.7, \kappa=.161$.
The four sets of points represent the different values of
RQM mass parameter $\mu$ used to construct smearing functions. One-state fits
were obtained from time-windows (reading from right to left on the graph) 1-6,
2-7, 3-8, 4-9. Points are plotted at $\exp[-(\Delta E)t_{min}]$, where $\Delta
E
=E_2 - E_1$ is the splitting of the first excited (2S) state from the
ground-state and $t_{min}$ is the smallest time included in the fit. The
2-state
fits were obtained from the window 2-8 but are plotted here at $t\approx
\infty$
to illustrate the convergence of the 1-state effective mass to the essentially
$\mu$ independent 2-state value.

\item Time-window dependence of ${\tilde f}_B$
for 1-state fit and comparison
with 2-state fit. (See item of Fig. 13.) Also plotted for comparison are
the values obtained
by extrapolating the 1-state fits to $t = \infty$, using the measured value of
the energy-splitting $E_2-E_1 = 0.32$.

\item Scales obtained from
$m_{\rho}$ (circles and filled circles), $f_{\pi}$ (squares), string tension
(diamonds and filled diamonds), and deconfinement temperature $T_c$
(filled squares) relative to the scales in Table 1. Our data is denoted
by open symbols.

\item $f_B$ at $\kappa=\kappa_c$ as a function of
lattice spacing. The
scales and renormalization constants used are given in Table 1. The solid line
is the best linear fit. The dashed line is a quadratic ($a^2$) fit used to
estimate systematic error in $a\rightarrow 0$ extrapolation. (See Section 5.5).

\item Ground state energy, in lattice units,
at $\kappa=\kappa_c$, versus
lattice spacing. Points at the same $a$ have been slightly displaced for
readability.

\item Mass splitting, $M_{B_s}-M_{B_u}$, in physical units,
versus lattice
spacing. Points at the same $a$ have been slightly displaced for readability.

\item Comparison of $f_{B_{s}}/f_{B_{u}}$ vs. lattice spacing for
present results and other recent works. Scale set from all
ACPMAPS values, using empirical fit,
$a^{-1}$ linear in $\beta$.  Points at the same
$a$ have been slightly displaced for readability.

\item Comparison of ${\tilde f}_B$ vs. lattice spacing for
present results and other recent works.
Points at the same $a$ have been slightly displaced for readability.

\item $m_{\pi}^2$ vs. naive quark mass for
$\beta = 5.7,5.9,6.1,$ and $6.3.$
The vertical scales for 5.9, 6.1, and 6.3 are offset by multiples of 0.2 for
display purposes. Points labeled 5.9e and 5.9u represent mesons with equal
and unequal quark masses, respectively. Values with unequal quark mass are
plotted at the average mass $(m_1+m_2)/2$.

\item Time evolution of the ground state
heavy-light wavefunction for
$\beta = 5.9, \kappa = .159$, using
a source smeared with the ground state wavefunction of the RQM with
constituent mass parameter $\mu = 0.12$.

\item Time evolution of the ground state
heavy-light wavefunction for
$\beta = 5.9, \kappa = .159$, using
a source smeared  over a cube of width 7 lattice spacings.

\item Time evolution of the first excited state
(2S) state heavy-light
wavefunction for $\beta = 5.9, \kappa = .159$, using
a source smeared over the 2S RQM wavefunction.

\item Time evolution of the first excited
(2S) state wavefunction using
a 3-state fit and selecting
the optimized combination of RQM smearing functions tuned to be orthogonal
to the ground state and second excited state.
a Bethe-Salpeter wavefunction and optimized smearing.

\end{list}


\begin{thebibliography}{99}
\bibitem{Eic81}
 E.~Eichten and F.~Feinberg, Phys. Rev. D {\bf 23}
 (1981) 2724.
 W.~E.~Caswell and G.~P.~Lepage, Phys. Lett. {\bf 234B} (1986) 437.
 E.~Eichten, in {\rm Field Theory on the Lattice\/},
 Nucl.~Phys.~B (Proc. Suppl.) {\bf 4} (1988) 170.
 G.~P.~Lepage and B.~A.~Thacker, ibid. 199.
\bibitem{Proof90}
 A proof that these conditions are satisfied for physical
 systems with one heavy quark has been given in leading
 order in $1/m_Q$ to all orders in the loop expansion by
 B.~Grinstein, Nucl. Phys. {\bf B339} (1990) 253.
\bibitem{EichtenHill1}
 E.~Eichten and B.~Hill, Phys. Lett. {\bf 234B} (1990) 511.
\bibitem{Georgi90}
 H.~Georgi, Phys. Lett. {\bf 240B} (1990) 447.
\bibitem{Wisgur}
 N.~Isgur and M.~B.~Wise, Phys. Lett {\bf 232B} (1989) 113;
 Phys. Lett {\bf 237B} (1990) 527.
\bibitem{Falk}
 A.~F.~Falk, H.~Georgi, B.~Grinstein and M.~B.~Wise, Nucl. Phys.
 B{\bf 343} (1990) 1.
\bibitem{Wise92} M.~B.~Wise, Proceedings of
the 1991 Lake Louise Winter Institute, p. 222, 1991(QCD161:L25:1991).
\bibitem{Bj}
 J.~D.~Bjorken, SLAC preprint SLAC-PUB-5278 (1990).
\bibitem{NR90}
 G.P.~Lepage and B.A.~Thacker, Phy. Rev D {\bf 43} (1991) 196.
\bibitem{Mandula} J.~Mandula and M.~Ogilvie, Nucl. Phys. (Proc. Suppl.)
{\bf 26} (1992) 459; Phys. Rev. D (Rap. Comm.){\bf 45} (1992) R2183.
\bibitem{KronMack} A.~X.~El-Khadra, A.~S.~Kronfeld, and P.~B.~Mackenzie,
FERMILAB-PUB-93/195-T.
\bibitem{BLS}C.~W.~Bernard, J.~N.~Labrenz, and A.~Soni, Phys. Rev. D{\bf 49}
(1994) 2536.
\bibitem{cubes} A. Bartolini et al, Nucl. Phys. B (Proc.Suppl.){\bf 30} (1993).
\bibitem{Wuppertal}C.~Alexandrou, S.~Gusken, F.~Jegerlehner, K.~Schilling, and
R.~Sommer,
CERN-TH 6692/92.
\bibitem{noise} C.~Bernard, C.~M.~Heard, J.~Labrenz and A.~Soni,
Nucl. Phys. B (Proc. Suppl.)
{\bf 26}, 384 (1992).
\bibitem{lat91_multistate}A.~Duncan, E.~Eichten, G.~Hockney, and H.~Thacker,
Nucl. Phys. B (Proc. Suppl.)
{\bf 26}, 391 (1992).
\bibitem{DET_RQM}A.~Duncan, E.~Eichten, and H.~Thacker, Phys.~Lett. B{\bf 303}
(1993) 109.
\bibitem{Rome2}C R Allton et al, Phys. Lett. B{\bf 326} (1994) 295.
\bibitem{UKQCD}R M Baxter et al, Phys. Rev. D{\bf 49} (1994) 1594.
\bibitem{LepMack}G.~P.~Lepage and
P.~B.~Mackenzie, Phys. Rev. D {\bf 48} (1993) 2250.
\bibitem{bernard}C.~Bernard, Nucl. Phys. B (Proc. Suppl.) {\bf 34} (1994) 47.
\bibitem{HernHill}O.~F.~Hern\'andez and B.~R.~Hill, Phys. Rev. {\bf D50}
(1994) 495.
\bibitem{polemass}S Narison, Phys Rep {\bf 84} (1982) 263 and Phys
Lett B {\bf 197} (1987) 405; E Bagan et al, Phys Lett B {\bf 278}
(1992) 457
\bibitem{pdg}Particle Data Group, Phys Rev {\bf D45} (1992) S1
\bibitem{polemassvalue}C A Dominguez and N Paver, Phys Lett B {\bf
293} (1992) 197; The NRQCD Collaboration, Nucl. Phys. B (Proc. Suppl.){\bf 34}
(1994) 417.
\bibitem{oneloopanomdim}M B Voloshin and M A Shifman, Yad Fiz {\bf 45}
 (1987) 463 [Sov J Nucl Phys {\bf 45} (1987) 292];
 H D Politzer and M B Wise, Phys Lett B {\bf 206} (1988) 681 and Phys
 Lett B {\bf 208} (1988) 504
\bibitem{JandM}X Ji and M J Musolf, Physics Letters B {\bf 257} (1991) 409
\bibitem{lat93_fb}A.~Duncan, E.~Eichten, J.M.~Flynn,
B.R.~Hill, and H.~Thacker, Nucl. Phys. B (Proc. Suppl.) {\bf 34},
444 (1994).
\bibitem{BandGone}D J Broadhurst and A G Grozin, Phys Lett B {\bf 267}
(1991) 105.
\bibitem{BandGtwo}D.~J Broadhurst and A G Grozin, Phys Lett B {\bf 274}
(1992) 421.
\bibitem{EichtenHillI}E.~Eichten and B.~Hill, Phys~Lett B {\bf 234}
(1990) 511.
\bibitem{FNL} A.F. Falk, M. Neubert, M. Luke
Nucl.Phys. B {\bf 388}(1992)363.
\bibitem{renormalons}I.I. Bigi, M.A. Shifman, N.G. Uraltsev, and
A.I. Vainshtein, ``The Pole Mass of the Heavy Quark. Perturbation Theory and
Beyond,'' hep-ph@xxx.lanl.gov - 9402360.
\bibitem{EichtenHillII}E.~Eichten and B.~R.~Hill, Phys.~Lett. B {\bf 240}
(1990) 193.
\bibitem{BorrelliAndPittori}A. Borrelli and C. Pittori,
Nucl. Phys. B {\bf 385} (1992) 502;
O.F. Hern\'andez and B.R. Hill,
Phys. Lett. B {\bf 289} (1992) 417.
\bibitem{Schnapka} E. Schnapka,``Validity of Relativistic Potential Models
in Confining Theories", Univ. of Pittsburgh Master's Thesis, Aug.1993.
\bibitem{cea}P.Cea, P. Colangelo, L.Cosmai, and G.Nardulli, Phys. Lett. B {\bf
206}
(1988) 691.
\bibitem{Gottlieb} S.~Gottlieb, J.~Kuti, D.~Touissaint, A.~Kennedy, S.~Meyer,
B.~Pendelton, and R.~Sugar, Phys. Rev. Lett. {\bf 55} (1985) 1958; N.~Christ
and A.~Terrano, Phys. Rev. Lett {\bf 56} (1985) 111.
\bibitem{Bali} G.~S.~Bali and K.~Schilling, Phys. Rev. D{\bf 47}, 661 (1993).
\bibitem{Weingarten} D.~Weingarten, Nucl. Phys.{\bf B34} (Proc. Suppl.), 29
(1994).
\bibitem{Rome}C.~R.~Allton, et al, Nucl Phys B {\bf 413} (1994) 461.
\bibitem{Boucaud}Ph.~Boucaud, C.~L.~Lin, and O.~Pene,
Phys.~Rev. D{\bf 40}, 1529 (1989) and Phys.~Rev.
D{\bf 41}, 3541 (1990)(E).
\bibitem{ElKhadra}A. X. El-Khadra, G. M. Hockney, A. S. Kronfeld, and P. B.
Mackenzie,
Phys.Rev.Lett. {\bf 69} (1992) 729.
\bibitem{RomeSoton}C R Allton et al, Nucl Phys B {\bf 349} (1991) 349
\bibitem{LuscherI}M.~Luscher,  in \underline{Progress in Gauge Field Theory},
G.~'tHooft, et al. (eds.)(Cargese, 1983). New York:Plenum 1984.
\bibitem{Hochberg}D.~Hochberg and H.~B.~Thacker, Nucl. Phys. {\bf B257}[FS14],
(1985) 729.
\bibitem{LuscherII}M.~Luscher, Comm.~Math.~Phys. {\bf 104} (1986) 177.
\bibitem{Fukugita}M.~Fukugita, H.~Mino, M.~Okawa, G.~Parisi, and A.~Ukawa,
Phys. Lett. {\bf B294} (1992) 380.
\bibitem{Kentucky}T Draper and C McNeile, Nucl. Phys. B (Proc. Suppl.) {\bf 34}
(1994) 453.
\bibitem{Hashimoto}S.~Hashimoto and Y.~Saeki, Mod.~Phys.~Lett. A7, 387 (1992).
\end{thebibliography}
\end{document}